\documentclass[useAMS,usenatbib]{mn2e}

\DeclareSymbolFont{cmletters}{OML}{cmm}{m}{it}
\DeclareMathSymbol{v}{\mathalpha}{cmletters}{"76}

\usepackage{tabularx,ragged2e,booktabs,caption}
\usepackage{amsmath}
\usepackage{amssymb}
\usepackage{epsfig}
\usepackage{graphicx}
\usepackage{ifthen}
\usepackage{latexsym}
\usepackage{rotating}
\usepackage{subfigure}
\usepackage{times,epsf}
\usepackage{txfonts}
\usepackage{varioref}
\usepackage{verbatim}
\usepackage{array}
\usepackage{url}
\usepackage{color}
\usepackage[dvipsnames]{xcolor}
\usepackage[T1]{fontenc}
\usepackage{ulem}

\newcolumntype{L}[1]{>{\raggedright\arraybackslash}p{#1}}
\newcolumntype{C}[1]{>{\centering\arraybackslash}p{#1}}
\newcolumntype{R}[1]{>{\raggedleft\arraybackslash}p{#1}}

\newcommand{\be}{\begin{equation}}
\newcommand{\ee}{\end{equation}}
\newcommand{\bea}{\begin{eqnarray}}
\newcommand{\eea}{\end{eqnarray}}

\newcommand\apj{Astrophysical Journal}
\newcommand\apjl{Astrophysical Journal Letters}

\newcommand\apjs{Astrophysical Journal Suppl. Ser.}

\newcommand\aap{Astronomy \& Astrophysics}

\newcommand\mnras{Monthly Notices of the Royal Astronomical Society}

\newcommand\pasj{Publications of the Astronomical Society of Japan}

\newcommand\ARAA{Ann. Rev. Astron. Asrophys.}
\newcommand\jqsrt{Journal of Quantitative Spectroscopy and Radiative Transfer}
\newcommand\araa{\ARAA}

\newcommand{\der}[2]{\frac{{\rm d}#1}{{\rm d}#2}}

\newcommand{\koral}{\texttt{KORAL}\,}
\newcommand{\Medd}{\dot M_{\rm Edd}}

\newcommand{\medd}{\dot M_{\rm Edd}}

\title[Energy flows in thick accretion disks and their consequences for black hole feedback]{Energy flows in thick accretion disks and their consequences for black hole feedback}
\author[A. S\k{a}dowski, J.-P. Lasota, M. A. Abramowicz, R. Narayan]
       {Aleksander S\k{a}dowski$^1$\footnotemark[1], 	 
         Jean-Pierre Lasota$^{2,3}$\footnotemark[1],
         Marek A. Abramowicz$^{3,4}$\footnotemark[1]     \newauthor
        and Ramesh Narayan$^{5}$\thanks{E-mail: asadowsk@mit.edu (AS); lasota@iap.fr (JPL); marek.abramowicz@physics.gu.se (MAA); rnarayan@cfa.harvard.edu (RN)} \\
        $^1$ MIT Kavli Institute for Astrophysics and Space Research
77 Massachusetts Ave, Cambridge, MA 02139, USA\\
$^2$ Institut d'Astrophysique de Paris, CNRS et Sorbonne
Universit\'es, UPMC Paris~06, UMR 7095, 98bis Bd Arago, 75014 Paris,
France\\
$^3$  Nicolaus Copernicus Astronomical Center, Bartycka 18, 00-716 Warsaw, Poland  \\
$^4$  Physics Department, Gothenburg University, SE-412-96 G\"oteborg, Sweden \\
$^5$ Harvard-Smithsonian Center for Astrophysics, 60 Garden St., Cambridge, MA 02134, USA}

\begin{document}

\maketitle

\label{firstpage}

\begin{abstract}
  We study energy flows in geometrically thick accretion discs, both
  optically thick and thin, using general relativistic,
  three-dimensional simulations of black hole accretion flows. We find
  that for non-rotating black holes the efficiency of the total
  feedback from thick accretion discs is $3\%$ - roughly half of the
  thin disc efficiency. This amount of energy is ultimately distributed between
  outflow and radiation, the latter scaling weakly with the accretion
  rate for super-critical accretion rates, and returned to the
  interstellar medium. Accretion on to rotating black holes is more
  efficient because of the additional extraction of rotational
  energy. However, the jet component is collimated and likely to
  interact only weakly with the environment, whereas the outflow and
  radiation components cover a wide solid angle.
\end{abstract}

\begin{keywords}
  accretion, accretion discs -- black hole physics -- relativistic
  processes -- methods: numerical
\end{keywords}

\section{Introduction}
\label{s.introduction}

Several aspects of the evolution of galaxies have been
puzzling astronomers for decades. Firstly, star formation in galaxies turns out
to be efficiently quenched in galactic bulges despite the gas cooling
time being much shorter than the age of a given galaxy \citep{cowie+77,fabiannulsen-77}. Secondly, the
galaxy luminosity function features a sharp high-mass cutoff
in which the most massive systems are red, dead and elliptical,
inconsistent with the hierarchical growth of structure in the Universe
\citep{croton+06}. Explaining both phenomena requires additional
processes preventing gas from collapsing into stars and limiting the
mass of the central galaxies.

Supernova explosions and stellar winds return energy (provide
\textit{feedback}) to the
interstellar medium (ISM). Although these processes take place at
small scales, they are powerful enough to affect the evolution of the
whole galaxy. Without strong stellar feedback, gas inside galaxies
would cool
efficiently and collapse on a dynamical time resulting in star
formation rates inconsistent  with observations. As shown recently 
shown by \citet{hopkins+14}, stellar feedback itself is enough to explain most of the
properties of galaxies, e.g., the relation between galaxy stellar mass
and halo mass, at stellar masses $M_*\lesssim 10^{11}M_\odot$.

Additional processes are needed to explain the formation of the most
massive galaxies. It is believed that almost  every galaxy harbours a supermassive black hole
(SMBH) in its nucleus. Being extremely compact such objects can
liberate gravitational energy in large amounts. As
a black hole (BH) grows to 0.2\% of the bulge mass through accreting matter, it releases
nearly 100 times the gravitational binding energy of its host galaxy
\citep{fabian+09}. It is therefore reasonable to expect that, if only
the energy returned from accretion (the
\textit{black hole feedback}) is efficiently coupled with the ISM, the
central SMBHs can strongly affect the formation and the properties of the host
galaxies. 

The feedback provided by SMBHs is therefore crucial for studying the
evolution of the Universe. It is often accounted for in large scale
simulations of galaxy formation, but the adopted models 
are very simplistic. The large range of scales involved in such
simulations does not allow for detailed numerical (and simultaneous)
modeling of the BH accretion. Instead, the mass supply
rate is estimated (at most) at parsec scales, usually using the Bondi
model of spherical accretion, and simple formulae for the feedback
efficiency are applied. These are based partly on the standard thin
disc models \citep{ss73}, but (to be consistent with observed
properties of galaxies) involve additional factors
arbitrarily rescaling the feedback rate.

These factors reflect our lack of understanding of how accretion on
SMBHs works and the efficiency of the feedback it provides. There are two major
unknown. Firstly, it is not clear how much matter
makes it to the BH, and how much is lost on the way. In other words,
what fraction of the gas attracted by the BH near the Bondi radius
ultimately crosses the BH horizon and efficiently liberates its
binding energy providing the energy source for the feedback \citep[see
discussion in][]{yuannarayan+14}. Secondly,
it is crucial to understand what fraction of this energy is returned
to the ISM.

In this paper we address the second question. The feedback efficiency
from an accretion flow is believed to be well established only for geometrically  thin
discs, corresponding to moderate, sub-Eddington accretion rates,
$10^{-3}\medd \lesssim \dot M\lesssim \medd$ (for the definition of
$\medd$ see Eq.~\ref{e.medd}). In this case, the accretion
flow is radiatively efficient, and all the released binding energy of the gas\footnote{The accretion itself is
not the only energy source in an accreting system. If the accretion
manages to bring a significant amount of magnetic flux on the BH,
magnetic jets can extract rotational energy of the BH. Jets, however,
are collimated, and may not interact efficiently with the ISM.} goes into
radiation and is determined by the binding energy of the  gas at the disc's
inner edge (e.g., it equals $5.7\%$ of the accreted rest mass energy,
$\dot M c^2$, for a non-rotating BH). 

We address here the question of what amount of energy is extracted if
accretion flows are not geometrically thin, i.e., how efficient the BH
feedback is if a SMBH accretes either in the radio mode ($\dot
M\lesssim 10^{-3}\medd$), when one
expects an optically  thin accretion flow and low radiative efficiency, or
above the Eddington accretion rate, in an optically thick disc. To
this purpose, we analyze a set of state-of-the-art,
three-dimensional simulations of the innermost region of BH accretion
performed with a general-relativistic,
radiative magnetohydrodynamical (MHD) code \koral \citep{sadowski+koral}.

Our paper has the following structure. In Section~\ref{s.thin} we
discuss the energy transfer in the standard model of a thin disc. In
Section~\ref{s.enflowinsims} we give the details of the numerical
simulations and dicuss their properties. In Section~\ref{s.discussion}
we discuss their implications and several caveats. Finally, in
Section~\ref{s.summary} we summarize our findings.

\section{Energy flow in thin discs}
\label{s.thin}

We start by recapitulating the physics of energy transfer in the standard model
of a thin accretion disc \citep[e.g.][]{ss73,frank+book}. This will give us a good reference point
when discussing energy flows in numerical simulations of accretion
flows. 

\subsection{Viscous dissipation}

The thin disc model assumes Keplerian azimuthal
motion, small vertical thickness of the disc, $h/r\ll 1$, and
radiative efficiency. Keplerian angular velocities imply differential
rotation and, in the presence of viscosity, non-zero transfer of
angular momentum between adjacent rings. The torque exerted by rings
on each other is \citep{frank+book},
\be
T=2\pi r\nu \Sigma r^2 \der{\Omega}{r}=-3\pi \nu \Sigma r^2 \Omega,
\label{e.torque}
\ee
where $\Sigma$ is the surface density at radius $r$,
$\Omega=\sqrt{GM/r^3}$ is the Keplerian
angular velocity, and $\nu$ is the local kinematic
viscosity coefficient corresponding to magnetically induced turbulence.

The torque results in transfer of angular momentum between the
rings. Conservation of angular momentum requires,
\be
\der{}{r}\dot M \Omega r^2=-\der{T}{r},
\ee
where $\dot M>0$ denotes the accretion rate.
Integrating between radius $r$ and the inner edge of the disc at
$r_{\rm in}$, we get,
\be
-\dot M \left(\Omega(r) r^2-\Omega(r_{\rm in})r_{\rm
    in}^2\right)=T(r)-T(r_{\rm in}).
\label{e.lcons}
\ee 
Following the standard assumption that the torque at the inner edge of
a thin disc vanishes \citep[see][]{paczynski-thindiscs} we get,
\be
T = -\sqrt{GM}\dot M \left(\sqrt{r} -\sqrt{r_{\rm in}}\right).
\ee

This torque not only transports angular momentum but also
dissipates mechanical energy heating up
the gas. The dissipation rate (per unit radius) in the whole ring which equals, by assumption of
radiative efficiency, minus the radiative cooling rate, is given by,
\be
q_{\rm diss}=-q_{\rm rad}=-T\der{\Omega}{r}=-\frac{3GM\dot M}{2r^2}\left(1-\sqrt{\frac{r_{\rm in}}{r}}\right),
\label{e.qvisc}
\ee
where the signs have been chosen such that a positive rate
corresponds to cooling, and negative to heating.
Dividing by the surface area of both sides of the ring we get
the well known thin-disc surface radiative flux,
\be
Q_{\rm rad}=\frac{q_{\rm rad}}{4\pi r}=\frac{3GM\dot M}{8\pi r^3}\left(1-\sqrt{\frac{r_{\rm in}}{r}}\right).
\label{e.Frad}
\ee
It is worth reiterating that the viscous dissipation rate (the
gas heating rate) does not depend on the particular form of the
viscosity (e.g., the $\alpha$-viscosity), but follows from the
assumptions of Keplerian motion, the zero-torque boundary condition and angular momentum conservation.

\subsection{Local energy budget}

In the previous paragraph we have shown that viscous dissipation in a
differentially rotating flow
results in heating of gas and radiative cooling at the rate 
$q_{\rm rad}$ (Eq.~\ref{e.qvisc}).

The energy required for this radiative emission may come from the gravitational
field -- gas approaching the BH liberates its own binding energy at the
rate,
\be
q_{\rm bind}=\dot M\der{e_{\rm
  bind}}{r}=\dot M\der{}{r}\left(-\frac{GM}{2r}\right)=\frac{GM\dot
  M}{2r^2}.
\label{e.qbind}
\ee
However, it is clear that,
\be
q_{\rm bind}+ q_{\rm rad}\neq 0.
\ee
This means that there must be another source or sink component in the
local budget of energy.

In the previous section we have seen that viscosity leads to the
transport of angular momentum and 
dissipation of mechanical energy. However, viscosity transports not only angular momentum but also
rotational energy. The amount of energy transported in this way is,
\be
L_{\rm visc}=-T\Omega,
\label{e.Lvisc1}
\ee
and the resulting local heating or cooling  rate per unit radius is given by,
\be
q_{\rm visc}=\der{}{r}(T\Omega)=\frac{GM\dot M}{r^2}\left(1-\frac32\sqrt{\frac{r_{\rm in}}{r}}\right).
\label{e.qviscvisc}
\ee
It is straightforward to verify that,
\be
q_{\rm bind}+ q_{\rm rad}+q_{\rm visc}=0.
\ee
The viscous energy transport redistributes energy released in the disc
and compensates for the imbalance between the local binding energy release
and the rate of radiative cooling.

\begin{figure}
\includegraphics[width=.95\columnwidth]{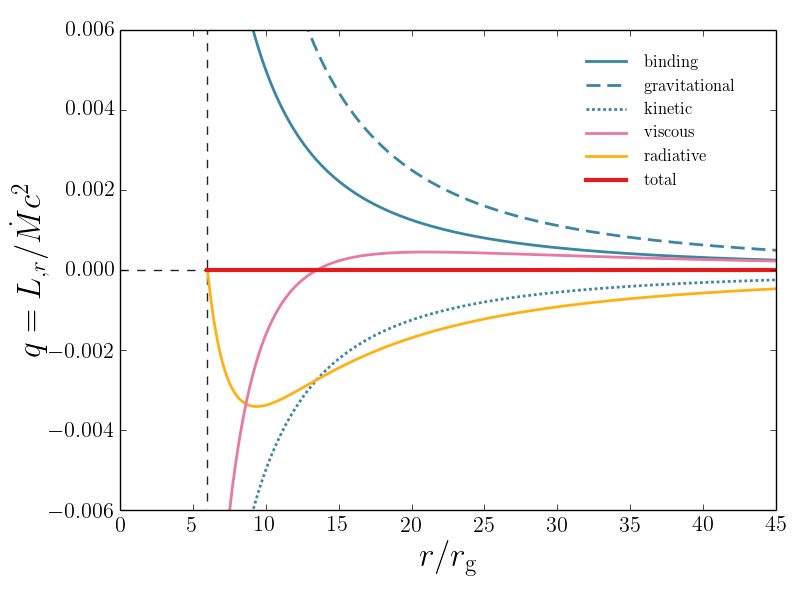}
\caption{Local energy gain in its various forms in  the standard thin
  disc model described in Section~\ref{s.thin}.}
\label{f.enfluxes_local_thin}
\end{figure}

In Fig.~\ref{f.enfluxes_local_thin} we plot local heating/cooling
rates in a thin disc as a function of radius. The solid blue line
shows the energy gain from the change in the binding energy, $q_{\rm
  bind}$ (Eq.~\ref{e.qbind}). This quantity is further
decomposed into the gravitational,
\be
q_{\rm grav}=\dot M\der{e_{\rm
  grav}}{r}=\dot M\der{}{r}\left(-\frac{GM}{r}\right)=\frac{GM\dot
  M}{r^2}.
\label{e.qgrav}
\ee
and kinetic,
\be
q_{\rm kin}=\dot M\der{e_{\rm
  kin}}{r}=\dot M\der{}{r}\left(\frac{GM}{2r}\right)=-\frac{GM\dot
  M}{2r^2}.
\label{e.qkin}
\ee
components. They are denoted by dashed and dotted blue lines,
respectively.

The orange line shows the radiative cooling rate, $q_{\rm rad}$
(Eq.~\ref{e.qvisc}). As expected, no emission comes from the inner edge
of the disc (located at $r_{\rm in}=6GM/c^2$, but we are using here the Newtonian approximation) and the most efficient emission takes place
from a ring located at $r \approx 8 r_{\rm g}$\footnote{The maximum is at $r = (49/6) \,r_{\rm g}$.}, where
$r_g=GM/c^2$. 

The pink line reflects the energy redistribution rate by the viscosity, $q_{\rm visc}$
(Eq.~\ref{e.qviscvisc}). For
$r\lesssim 13 r_{\rm g}$ it is negative -- at these radii viscosity effectively
cools the disc and carries the energy outward. It is particularly
evident for the gas approaching the inner edge, where the release of
binding energy is large, but the no-torque boundary condition prevents
radiative emission. To maintain the energy balance, viscosity must
transport this locally liberated binding energy out.

For $r\gtrsim
13 r_{\rm g}$, $q_{\rm visc}$ becomes positive, which means that the
viscous energy flux decreases with increasing radius and locally deposits
energy, contributing to the local heating rate and increasing the
magnitude of radiative cooling beyond the rate at
which local binding energy is liberated. In the limit $r\gg r_{\rm
  in}$ one has $q_{\rm rad}=-3q_{\rm bind}$, i.e., the local
rate of releasing energy in radiation is three times larger than the
change in binding energy. The extra contribution comes from the
viscous energy flux which deposit energy (and heats up gas) at a rate two times
larger than the gain from released binding energy.

\subsection{Energy flow}

In the previous section we have looked into the local energy
balance. Now, let us look into the total amount of energy carried by
its various components from one radius to another.

The binding energy is carried by the flow at a rate,
\be
L_{\rm bind}=-\dot M e_{\rm bind}=\frac{GM\dot M}{2r}>0,
\label{e.Lbindthin}
\ee
where the positive sign reflects the fact that bound gas is falling
inward, thus effectively depositing energy at infinity. The luminosity in
binding energy may be again decomposed into the gravitational and
kinetic components,
\be
L_{\rm grav}=-\dot M e_{\rm grav}=\frac{GM\dot M}{r}>0,
\label{e.Lgravthin}
\ee
\be
L_{\rm kin}=-\dot M e_{\rm kin}=-\frac{GM\dot M}{2r}<0.
\label{e.Lkinthin}
\ee
Gravitational energy luminosity is positive, but the kinetic
luminosity is negative -- kinetic energy of the Keplerian motion is
brought inward by the gas.

The radiative cooling rate, $q_{\rm rad}$, is given by
Eq.~\ref{e.qvisc}. Photons are emitted from the disc surface and
leave the system. The total radiative luminosity at given radius,
$L_{\rm rad}$, 
results from the emission inside that radius,
\be
L_{\rm rad}=\int_{r_{\rm in}}^r q_{\rm rad} \,{\rm dr} = \frac32
\frac{GM\dot M}{r}\left(\frac13\frac {r}{r_{\rm in}} + \frac
  23\sqrt{\frac{r_{\rm in}}{r}}-1\right)>0.
\label{e.Lradthin}
\ee
This quantity is zero at the inner edge ($r=r_{\rm in}$) and equals
$GM\dot M/2r_{\rm in}$ at infinity. The radiative luminosity of the
whole accretion disc is therefore equal to the binding energy of the
gas crossing the inner edge.

Finally, the amount of energy carried by viscosity from the inner region outward, $L_{\rm visc}$, is
(Eq.~\ref{e.Lvisc1}),
\be
L_{\rm visc}=-T\Omega=\frac{GM\dot M}{r}\left(1-\sqrt{\frac{r_{\rm in}}{r}}\right)>0.
\label{e.Lvdiscthin}
\ee

\begin{figure}
\includegraphics[width=.95\columnwidth]{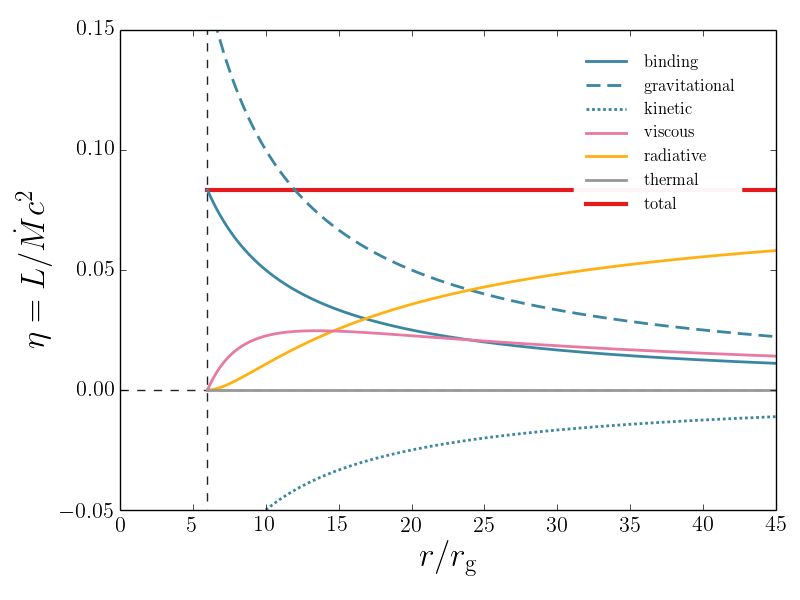}
\caption{Luminosity in various forms of energy for the standard thin
  disc model described in Section~\ref{s.thin}. The thick red line
  denotes the total luminosity of the system which can be decomposed
  into the luminosity in binding energy (solid blue line), in
  radiation (orange) and luminosity transported by viscosity
  (pink). The luminosity in binding energy is further decomposed into
  gravitational (blue dashed) and kinetic (blue dotted)
  components. All the luminosities are normalized with the accreted
  rest-mass energy, $\dot M c^2$.}
\label{f.enfluxes_thin}
\end{figure}

The various integrated energy fluxes introduced above are shown in
Fig.~\ref{f.enfluxes_thin}. Their magnitudes have been normalized to the
amount of accreted rest-mass energy. The blue lines show the
luminosities in the binding energy, $L_{\rm bind}$, and its
gravitational and kinetic components,
$L_{\rm grav}$ and $L_{\rm kin}$. At the inner edge ($r_{\rm in}=6r_g$), the amount of
binding energy carried by the gas is 
\be
L_{\rm bind,in}=\frac{GM\dot M}{2r_{\rm
    in}}=\frac{1}{12}{\dot M c^2},
\ee
which is, as we will discuss in detail in a moment, the total
efficiency of a thin disc in the Newtonian gravitational potential.

The orange line in Fig.~\ref{f.enfluxes_thin} shows the radiative
luminosity crossing a sphere of radius $r$. As no photons are emitted
from inside the inner edge, it starts from zero and
gradually grows, reaching finally $GM\dot M/2r_{\rm
  in}$ at infinity -- in a thin disc, the whole energy extracted by the infalling
gas ultimately goes into radiation. 

It is interesting to note that
$50\%$ of the radiation is emitted from outside $r\approx 25r_{\rm g}$. At the same
time, the gas infalling from infinity down to that radius has
extracted only roughly $25\%$ of the available binding energy. The
excess in radiative luminosity comes from the extra energy carried by
viscosity from the innermost region.

This component of the energy flux is denoted with the pink line at the same plot. The
amount of energy carried by viscosity grows rapidly just outside of
the inner edge -- at these radii
viscosity is transporting rotational kinetic energy outward. Outside
$r\approx 13r_{\rm g}$ the luminosity of viscous energy transport
drops down with radius and the energy taken away from the innermost
region is deposited by viscosity into the gas.

Summing up all the components of the energy transfer we get the total
luminosity,
\be
L_{\rm tot}=L_{\rm bind}+L_{\rm rad}+L_{\rm visc},
\label{e.Ltotthin}
\ee
which is the quantity that is fundamentally conserved in stationary
flows, i.e., is independent of radius and no energy accumulates at any
location. Indeed, the sum of the three components (red line in
Fig.~\ref{f.enfluxes_thin}) gives a constant value equal to the total
efficiency of accretion and the binding energy carried in by the gas
through the disc inner edge ($1/12)\dot M c^2$. In the Schwarzschild metric this
efficiency would be $\sim 0.057\dot M c^2$.

\section{Energy flow in simulations of accretion flows}
\label{s.enflowinsims}

Having recapitulated how energy flows in a standard thin
disc, we are ready to study the energy redistribution in numerical simulations of accretion flows.
In the following Section we describe the numerical method used to
perform the simulations. In Section~\ref{s.energyfluxes} we introduce the
formalism used to study energy fluxes in numerical solutions. In Sections~\ref{s.adafs} and \ref{s.slim} we look in detail
into the energy flow in simulations of optically thin and thick discs,
respectively.

\subsection{Numerical setup}
\label{s.numerical}

The simulations analyzed in this paper were performed in three
dimensions with the general relativistic radiation magnetohydrodynamical (GRRMHD) code
\texttt{KORAL} \citep{sadowski+koral} which solves the conservation
equations in 
a fixed, arbitrary spacetime using finite-difference methods. The
equations we solve are,
\bea\label{eq.rhocons}
\hspace{1in}(\rho u^\mu)_{;\mu}&=&0,\\\label{eq.tmunucons}
\hspace{1in}(T^\mu_\nu)_{;\mu}&=&G_\nu,\\\label{eq.rmunucons}
\hspace{1in}(R^\mu_\nu)_{;\mu}&=&-G_\nu,
\eea
where $\rho$ is the gas
density in the comoving fluid frame, $u^\mu$ are the components of the gas four-velocity, $T^\mu_\nu$ is the
MHD stress-energy tensor,
\be\label{eq.tmunu}
T^\mu_\nu = (\rho+u_{\rm g}+p_{\rm g}+b^2)u^\mu u_\nu + (p_{\rm g}+\frac12b^2)\delta^\mu_\nu-b^\mu b_\nu,
\ee 
$R^\mu_\nu$ is the stress-energy tensor of radiation, and $G_\nu$ is the radiative
four-force describing the interaction between gas and radiation \citep[see][for a more detailed description]{sadowski+koral2}. Here, $u_{\rm g}$ and $p_{\rm g}=(\Gamma-1)u_{\rm g}$ represent the internal energy and pressure of the 
gas in the comoving frame and $b^\mu$ is the magnetic field 4-vector \citep{gammie03}.
The magnetic pressure is $p_{\rm mag}=b^2/2$ in geometrical units. 

The magnetic field is evolved via the induction equation,
\be
\label{eq.Maxi}
\partial_t(\sqrt{-g}B^i)=-\partial_j\left(\sqrt{-g}(b^ju^i-b^iu^j)\right),
\ee
where $B^i$ is the magnetic field three-vector \citep{komissarov-99},
and $\sqrt{-g}$ is the metric determinant.
The divergence-free criterion is enforced using the flux-constrained 
scheme of \cite{toth-00}. 

The radiation field is evolved
through its energy density and flux, and the radiation stress-energy
tensor is closed by means of the M1 closure scheme
\citep{levermore84,sadowski+koral}. The energy exchange between gas
and radiation is by free-free emission/absorption as well as Compton scattering.
The latter is treated in the ``blackbody''
Comptonization approximation as described in \citet{sadowski+comptonization}.

We use modified Kerr-Shild coordinates with the inner edge of the
domain inside the BH horizon. The simulations are run with a
moderately high resolution of 252 grid cells spaced logarithmically in
radius, 234 grid cells in the polar angle, concentrated towards the
equatorial plane, and 128 cells in azimuth.

Three of the four simulations which we analyze in this work are
identical to the ones presented in \cite{sadowski+3d}. To have a
consistent optically thin version of an accretion flow we simulated an
additional model
with purely magnetohydrodynamical evolution, i.e., without radiation
field. This simulation (\texttt{h001}) corresponds to an optically thin,
advection dominated accretion flows (ADAF) believed to occur in
systems accreting well below the Eddington level \citep{yuannarayan+14}.

Parameters of the models are given in Table~\ref{t.models}.

In this work we adopt the following definition
for the Eddington mass accretion rate,
\be
\label{e.medd}
\Medd = \frac{L_{\rm Edd}}{\eta c^2},
\ee
where $L_{\rm Edd}=1.25 \times 10^{38}  M/M_{\odot}\,\rm ergs/s$ is the 
Eddington luminosity, and $\eta$ is the radiative efficiency of a thin
disc around a black hole with a given spin $a_* \equiv a/M$. For zero BH spin,
$\Medd = 2.48 \times 10^{18}M/M_{\odot}  \,\rm g/s$.
Hereafter, we also use the
gravitational radius $r_{\rm g}=GM/c^2$ as the unit of length, and $r_g/c$
as the unit of time.

In this study we consider simulation
output averaged over time. Therefore, whenever we write, e.g., $\rho u^r$, we mean the
average of the product, i.e., $\langle \rho u^r \rangle$, where
$\langle \rangle$ stands for time averaging.

\begin{table}
\centering
\caption{Model parameters}
\label{t.models}

\begin{tabular}{lcccc}
\hline
\hline
&  \texttt{h001} & \texttt{r001} & \texttt{r003} & \texttt{r011}\\
\hline
& hydro & radiative & radiative & radiative\\
\hline
$M_{\rm BH}$ & $10 M_\odot$& $10 M_\odot$& $10 M_\odot$& $10 M_\odot$ \\
$\dot M/\Medd$  &   $\lesssim 10^{-3}$ & 10.0  & 175.8 & 17.4 \\
$a_*$ &   0.0 &   0.0 &   0.0 & 0.7 \\
$t_{\rm max}$ &  23,000 & 20,000 & 19,000 & 16,100 \\
\hline
\hline

\multicolumn{5}{l}{All models initiated as in \cite{sadowski+3d}.}\\
\multicolumn{5}{l}{$M_{\rm BH}$ - mass of the BH, $\dot M$ - average
  accretion rate, }\\
\multicolumn{5}{l}{$a_*$ - nondimensional spin parameter,}\\
\multicolumn{5}{l}{$t_{\rm max}$ - duration of the simulation in units of $GM/c^3$ }
\end{tabular}
\end{table}

\subsection{Energy fluxes}
\label{s.energyfluxes}

\subsubsection{Fundamental quantities}
\label{s.fundamental}

In quasi-stationary state the accretion rate is constant in radius,
i.e., gas does not accumulate anywhere, but rather 
flows towards the BH with constant rate. The accretion rate (the luminosity in rest-mass energy) is given by,
\be \label{e.mdot} 
\dot M = \int_{0}^\pi
\int_0^{2\pi}\sqrt{-g}\,\rho u^r{\rm d}\phi {\rm d}\theta,
\ee
where this and the following integrals are evaluated at a fixed
radius $r$.

In a similar way we may define the luminosity in all forms of energy,
\be \label{e.entot0} 
L_{\rm tot,0} = -\int_{0}^\pi
\int_0^{2\pi}\sqrt{-g}\,(T^r_t + R^r_t){\rm d}\phi {\rm d}\theta,
\ee
where we integrate the radial flux of energy carried by gas ($T^r_t$)
and by radiation ($R^r_t$). This quantity, however, is not interesting
from the point of view of a distant observer. It contains the flux of
rest-mass energy which, even if deposited at infinity, will not have
observational consequences
(since at infinity rest-mass cannot be converted into other
forms of energy in a trivial way). Therefore, we define the \textit
{total luminosity} 
 by subtracting\footnote{Lower time index
  introduces a negative sign in $T^r_t$, so to get rid of the
  rest-mass component in $T^r_t$ one has to \textit{add} $\rho u^r$.} the rest-mass energy flux from the
previous definition,
\be \label{e.entot} 
L_{\rm tot} = -\int_{0}^\pi
\int_0^{2\pi}\sqrt{-g}\,(T^r_t + R^r_t+\rho u^r){\rm d}\phi {\rm d}\theta,
\ee
The sign has been chosen in such a way that $L_{\rm tot}$ is negative for
energy falling in the BH, and positive for energy leaving the
system. 

In a stationary state the total 
  luminosity is independent of radius
  (if it was  not, energy would accumulate in some
  regions). It is the
luminosity of the whole system, i.e., it is also the luminosity
as seen from infinity. Therefore, it determines the rate at which energy is
deposited in the interstellar medium or, in other words,  $L_{\rm
  tot}$ is the total power of \textit{feedback}.

\subsubsection{Decomposition}
\label{s.decomposition}

The total energy flux consists of multiple components. We decompose it
in a way which gives well-known Newtonian limits. 

First, we single out
the radiative component and define the radiative luminosity,
\be \label{e.enrad} 
L_{\rm rad} = -\int_{0}^\pi
\int_0^{2\pi}\sqrt{-g}\,R^r_t{\rm d}\phi {\rm d}\theta,
\ee
which reflects energy carried by photons, either trapped in the
gas, or propagating freely.
To define other components, let us first write explicitly,
\be
\label{e.Trt}
T^r_t+\rho u^r=\rho u^r(1+u_t)+(\Gamma u_{\rm g}+b^2)u^r u_t -b^r b_t.
\ee 
Here we remind the reader that in all the integrals we take averages of products,
e.g., $\rho u^r(1+u_t)$ is actually $\langle\rho u^r(1+u_t)\rangle$,
where the product is averaged over time. This particular
quantity is the average radial flux of binding energy. In detail, it is
the sum of advective, $\langle \rho u^r\rangle\langle
1+u_t\rangle$, and Reynolds (turbulent), $\langle \rho u^r(
1+u_t)\rangle-\langle \rho u^r\rangle\langle
1+u_t\rangle$, components. Similar decomposition applies to the other
components of the total energy flux. In this work we will not
discriminate between the turbulent and advective fluxes, but instead
focus on the net contribution.

It is straightforward to define the luminosity in internal (thermal)
energy,
\be \label{e.enint} 
L_{\rm int} = -\int_{0}^\pi
\int_0^{2\pi}\sqrt{-g}\,\Gamma u_{\rm g} u^r u_t {\rm d}\phi {\rm d}\theta,
\ee
which, similarly, contains the advective and convective terms, and the luminosity
carried by the magnetic field,
\be \label{e.enmagn} 
L_{\rm magn} = -\int_{0}^\pi
\int_0^{2\pi}\sqrt{-g}\, (b^2 u^r u_t -b^r b_t).{\rm d}\phi {\rm d}\theta.
\ee
 which again includes both the advective
component and 
turbulent stress.
The remaining term (proportional to $(1+u_t)$) contains information about the gravitational and
kinetic energies. In the Newtonian limit it gives $-1/2r$ for
Keplerian motion. Therefore, we identify the corresponding integrated
energy flux as the luminosity in binding energy carried radially by gas,
\be \label{e.enbind} 
L_{\rm bind} = -\int_{0}^\pi
\int_0^{2\pi}\sqrt{-g}\,\rho u^r(1+ u_t) {\rm d}\phi {\rm d}\theta.
\ee
The gravitational component of the last expression can be singled out
by calculating the specific binding energy $(1+ u_t)$ for a stationary
observer. From $u^\mu=(u^t,\vec 0)$ and $u^\mu u_\mu=-1$ one gets (for
a diagonal metric)
$u_t=-\sqrt{-g_{tt}}$, and therefore, the luminosity in gravitational
energy carried by gas is,
\be \label{e.engrav} 
L_{\rm grav} = -\int_{0}^\pi
\int_0^{2\pi}\sqrt{-g}\,\rho u^r(1-\sqrt{-g_{tt}}) {\rm d}\phi {\rm d}\theta.
\ee
The remaining term reflects the luminosity in kinetic energy,
\be \label{e.enkin} 
L_{\rm kin} = L_{\rm bind} - L_{\rm grav}.
\ee

To sum up, we have decomposed the total energy transfer rate into
binding, thermal, magnetic, and radiative components,
\be
L_{\rm tot} = L_{\rm bind} + L_{\rm int} + L_{\rm magn} + L_{\rm rad}.
\ee

\subsubsection{Advective and viscous energy fluxes}
\label{s.magnetic}

In a viscous accretion flow energy is transported both by viscosity
and by the fluid which advectively carries energy with itself. One may
write,
$L_{\rm hydro}=L_{\rm adv}+L_{\rm visc}=\dot M Be - T\Omega$, where $Be$ is the
Bernoulli function of the fluid (which is not constant, because work
is done on gas on its way towards the BH), and $T\Omega$ reflects the viscous rate
of energy flow (Eq.~\ref{e.Lvdiscthin}). The hydrodynamical quantities
defined in the previous section ($L_{\rm bind}$, $L_{\rm int}$,
$L_{\rm magn}$) are based on time averaged quantities. The turbulence,
which provides effective viscosity, is averaged out and contributes to
the energy transfer rate. Therefore, as stated in the previous
Section, these luminosities include both terms,
the advective and the viscous one. It is beyond the scope of this
paper to decompose
them and single out the energy transfer rate solely due to viscosity.

\begin{figure}
\includegraphics[width=.95\columnwidth]{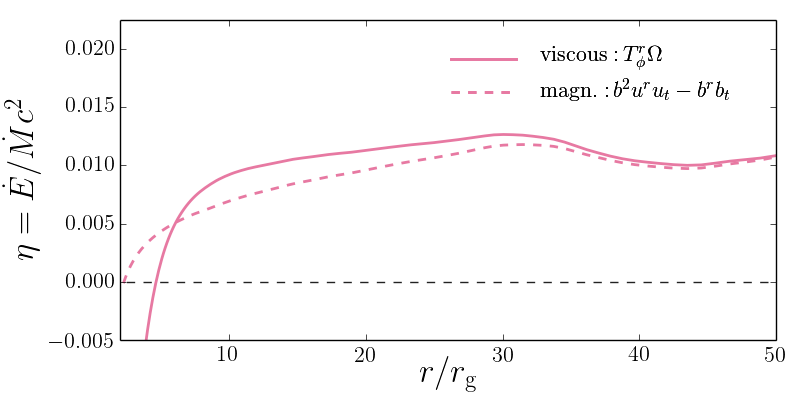}
\caption{Total estimated viscous flux of energy (solid line), and the
  energy carried by magnetic fields, $L_{\rm magn}$
  (Eq.~\ref{e.enmagn}, dashed line).}
\label{f.viscvsmagn_h001}
\end{figure}

We estimated the viscous component by calculating\footnote{If the
  viscous stress is proportional to shear then it is orthogonal to the
  gas velocity, $T_{\rm visc,\nu}^\mu u^\nu=0$. For purely
azimuthal motion, $u^\mu=(u^t,0,0,u^\phi)$, one finds that $T^r_\phi
\Omega=-T^r_t$. Therefore, $T^r_\phi\Omega$ indeed gives the radial
flux of energy carried by viscosity. However, in the simulations we
performed, the orthogonality and perfectly circular motion are not
enforced and these conditions are only approximately satisfied.}.
\be
L_{\rm visc,est} = -\int_{0}^\pi
\int_0^{2\pi}\sqrt{-g}\,(T^r_\phi \Omega){\rm d}\phi {\rm d}\theta.
\ee
This quantity is plotted in Fig.~\ref{f.viscvsmagn_h001} with the pink
solid line. In the same figure, we plot the magnetic component of the
luminosity, $L_{\rm magn}$ (Eq.~\ref{e.enmagn}). The two have very
similar profiles and magnitudes. This should not be surprising,
because it is mostly magnetic field which mediates angular
momentum transfer \citep[local shearing sheet and global
simulations of magnetized accretion show that the magnetic stress
dominates over the Reynolds stress by a factor of $\sim4$, see][]{pessah+06,penna+alpha}. From now
on, we will consider the luminosity in the magnetic component, $L_{\rm
  magn}$, as the counterpart of the viscous luminosity $L_{\rm
  visc}$
introduced in Section~\ref{s.thin}. Such assignment is helpful, but not
crucial, for the following considerations.

Often in the literature \citep[e.g.][]{abramowicz+88,narayanyi-94} the energy balance is
written in the comoving frame in the following form,
\be
\widehat q^{\rm heating}-\widehat q^{\rm cooling}=\widehat q^{\rm adv},
\ee
where $\widehat q^{\rm heating}$ and $\widehat q^{\rm cooling}$ stand
for local comoving heating
and cooling rates, and $\widehat q^{\rm adv}$ decribes the net amount of heat taken
away with the fluid or effectively brought in and locally
released. This particular decomposition
is not very helpful for the present study. However, we note that
for both the optically thin and thick discs, as will be discussed
below, the power advected with
the fluid (in thermal and radiative energies, respectively)
dominates the energy balance. Therefore, the flows discussed below
are indeed advection dominated.

\subsection{Energy flow in optically thin ADAFs}
\label{s.adafs}

Let us now look at the energy flow in simulated, multi-dimensional
accretion flows. We start with an optically thin disc (ADAF), model
\texttt{h001}). 

According to the standard model
\citep{narayanyi-94,abramowicz+adafs} for this mode of
accretion, energy locally dissipated does not have a chance to escape
because of low radiative efficiency 
and is advected with the flow. This fact makes such discs very hot and
geometrically thick. As a result, the expected efficiency of accretion
is zero because all the binding energy gained by gas on its way towards the
BH is balanced by thermal energy advected with it on the BH.

This model, however, does not allow the gas to flow out of the
system. This process, in principle, can provide a path for the
liberated binding energy to escape from the system, and as a result
may increase the efficiency of accretion.

\subsubsection{Luminosities}

\begin{figure}
\includegraphics[width=.95\columnwidth]{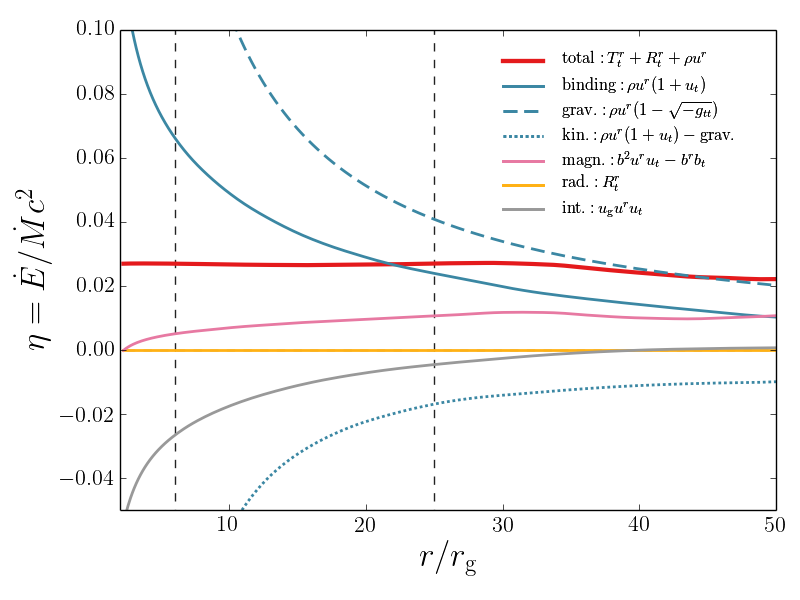}
\caption{Similar to Fig.~\ref{f.enfluxes_thin} but for a GRMHD
  simulation of an optically thin disc (ADAF, model \texttt{h001}). Colors denote the same
  components of luminosity. Additional gray line reflects the
  luminosity in thermal (internal) energy. Zero BH spin was
  assumed. For definitions see Section~\ref{s.adafs}. Vertical lines
  denote the ISCO and the estimated convergence region of the simulation
  at $r\approx 25$.}  
\label{f.enfluxes_h001}
\end{figure}

Figure~\ref{f.enfluxes_h001} presents the integrated radial fluxes
(luminosities) of
energy in various forms for the optically thin simulation
\texttt{h001}. The amount of binding energy (Eq.~\ref{e.enbind}) carried with the flow is
shown with solid blue line. The closer the gas gets to the BH, the
more bound it is, and the more luminosity it extracts with respect to
infinity (once again, infalling bound gas effectively deposits energy
at infinity). It can be decomposed into the gravitational 
(Eq.~\ref{e.engrav}, blue dashed line) and the kinetic
(Eq.~\ref{e.enkin}, blue dotted line) components. Because the flow is
only slightly sub-Keplerian 
and the radial velocities involved are low,
these two components behave qualitatively in the same way as in the
case of the thin disc discussed in the previous Section.

The magnetic component (Eq.~\ref{e.enmagn}), which reflects the energy
carried by effective viscosity, also qualitatively agrees with the
thin disc prediction. It is zero at the inner edge (which is now at
the horizon, not at innermost stable circular orbit (ISCO), because for thick discs stress is not zero
down to the horizon), and becomes positive, which again reflects
the fact that turbulent viscosity takes energy out of the innermost
region (here from $r\lesssim 10$) and carries it outward. In contrast
to the thin disc model, however, there is no clear decrease in the
magnetic luminosity inside the convergence region of the simulation,
i.e., turbulent viscosity does not contribute there to the local
heating rate.

Because radiative cooling is not efficient, energy is not transfered
by radiation. The dissipated energy is trapped in the flow, heats up
the gas, and contributes to the thermal energy transport
(Eq.~\ref{e.enint}). This fact is reflected in the grey line profile
in Fig.~\ref{f.enfluxes_h001}. The luminosity in thermal energy is no
longer negligible, as it was in the thin disc case. Significant amount
of thermal energy is carried inward with the flow. Because gas becomes
hotter when approaching the BH horizon, the corresponding magnitude
increases.

If the studied accretion flow followed the standard model, i.e., all
the energy released was advected on to the BH, all the components
contributing to the energy transfer should sum up to zero total
efficiency. This is, however, not the case. The thick red line in
Fig.~\ref{f.enfluxes_h001} reflects the total luminosity defined
through Eq.~\ref{e.entot}, i.e., composed of binding, magnetic and
thermal components. It is flat to a good accuracy inside $r\approx 25$
proving that the flow has reached a quasi-stationary state in this
region. The efficiency of $\sim3 \% \dot Mc^2$, reflects the amount of
energy extracted from the accretion flow\footnote{This value gives the
  total energy of feedback, in contrast to values given in
  \cite{sadowski+outflows} who gave only power in jet and wind components.}, and equals roughly 50\% of the
thin disc efficiency.

In the ideal case of an optically thin disc extending to infinity, this
amount of energy would be deposited at infinity. In practice, this is
the amount by which the BH systems affects the ISM (the BH feedback). Because
simulations with inefficient radiative cooling are scale-free, this
efficiency is characteristic of an optically thin flow (ADAF) at
\textit{any} accretion rate for which such a solution exists with negligible
radiative cooling \citep[i.e., for $\dot M\lesssim
10^{-3}\dot M$, see][]{yuannarayan+14}. The fate of the energy coming
out of the innermost region is discussed below in Section~\ref{s.fate}.

\subsubsection{Angular distribution}

\begin{figure}
  \rotatebox{90}{\hspace{1.2cm}Density}\hspace{.05cm}\includegraphics[width=1.0\columnwidth]{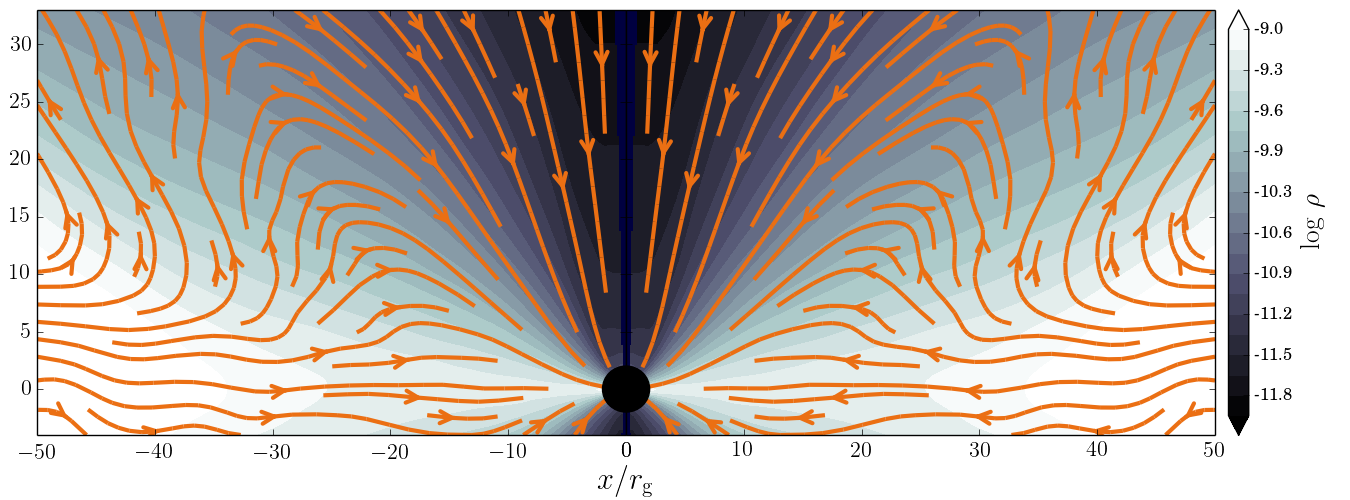}\vspace{-.3cm}
  \rotatebox{90}{\hspace{1.3cm}Rest mass}\hspace{.05cm}\includegraphics[width=1.0\columnwidth]{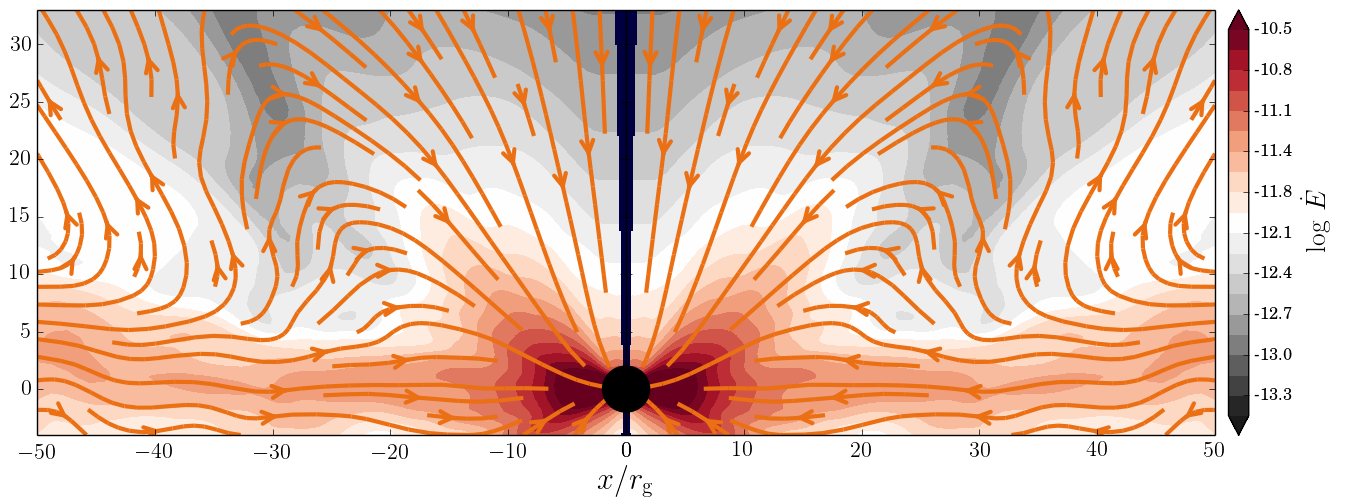}\vspace{-.3cm}
  \rotatebox{90}{\hspace{1.4cm}Total}\hspace{.05cm}\includegraphics[width=1.0\columnwidth]{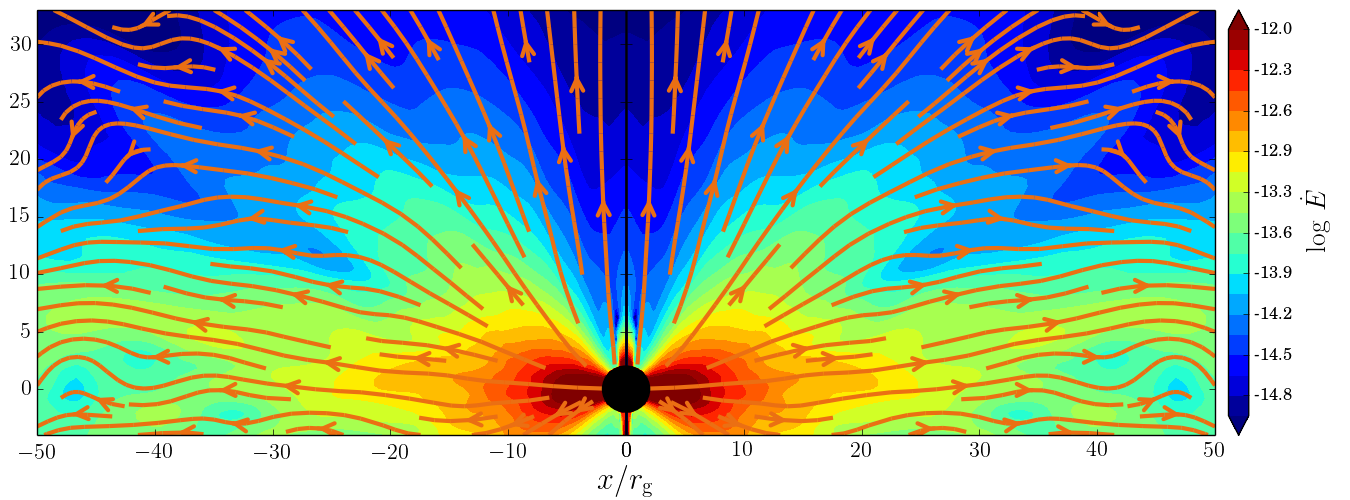}\vspace{-.3cm}
\rotatebox{90}{\hspace{1.3cm}Binding}\hspace{.05cm}\includegraphics[width=1.0\columnwidth]{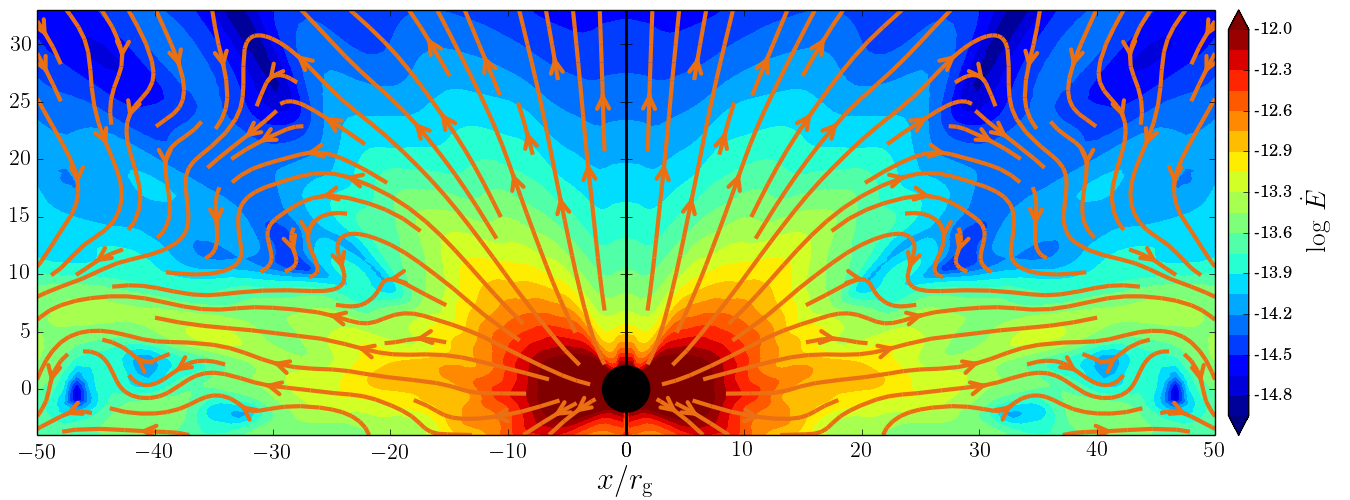}\vspace{-.3cm}
\rotatebox{90}{\hspace{1.cm}Magnetic/Visc.}\hspace{.05cm}\includegraphics[width=1.0\columnwidth]{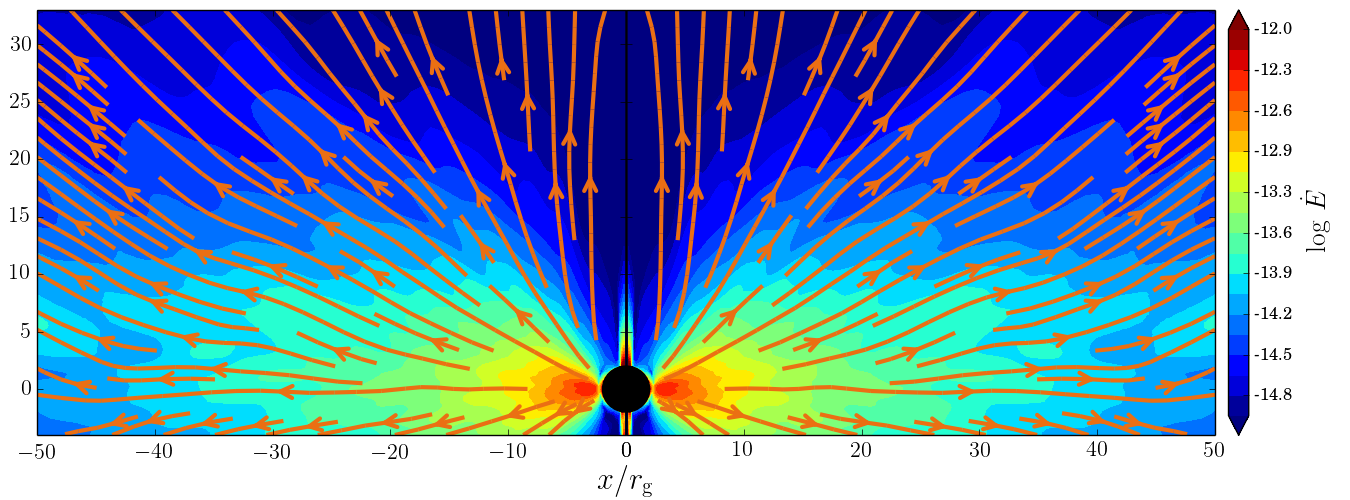}\vspace{-.3cm}
\rotatebox{90}{\hspace{1.3cm}Thermal}\hspace{.05cm}\includegraphics[width=1.0\columnwidth]{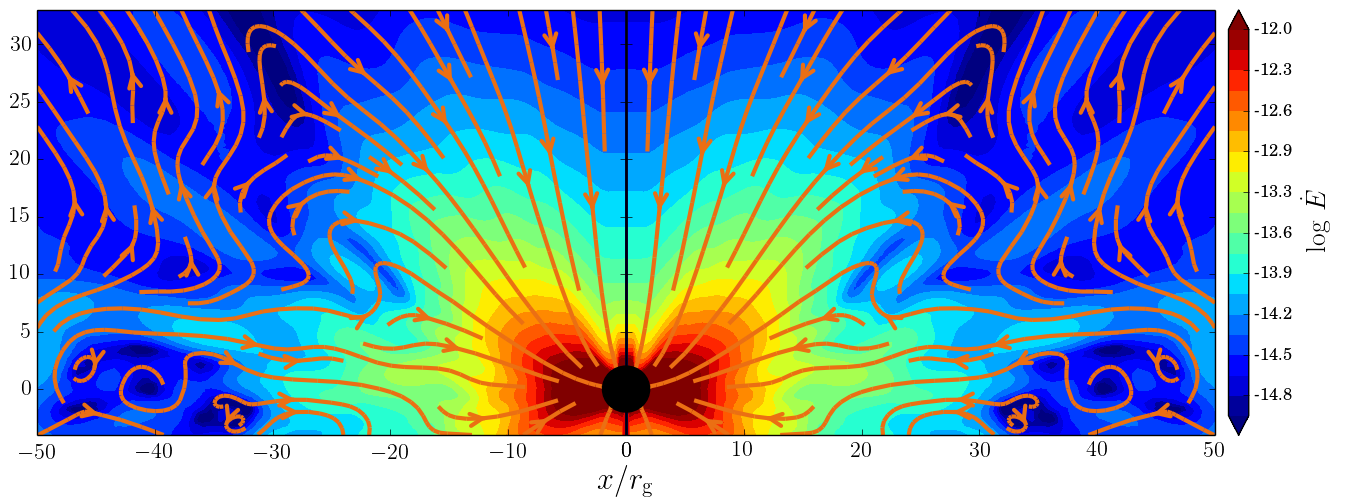}
\caption{Top panel: Distribution of averaged gas density in optically
  thin disc (model
  \texttt{h001}). Streamlines reflect direction of average gas
  velocity. Second panel: Magnitude of the rest mass density flux
  (colors) and its direction (streamlines). Third to sixth panels:
  Magnitudes and directions of energy fluxes (total, binding, viscous
  and Thermal, respectively).}
\label{f.sims_h001}
\end{figure}

Figure~\ref{f.sims_h001} shows the spatial distribution of density
(top-most panel) and various
components of the energy flux (other panels) in the optically thin simulation
\texttt{h001}. The streamlines in the top-most panel reflect the
velocity of the gas. The second panel shows the corresponding rest-mass
energy flux, $\rho u^\mu$. Most of the accretion takes place near the
equatorial plane. Within radius $r=30$, gas at all polar angles falls
inward. Only outside this radius (and outside the converged region of
the simulation), there is a hint of outflows that may arise from the
accretion flow. 

The third panel shows the magnitude (colors) and 
direction on the poloidal plane of the total energy flux, $L_{\rm
  tot}$ (Eq.~\ref{e.entot}). The total energy flows outward, in
agreement with the positive total efficiency of $3\%$ (see
Fig.~\ref{f.enfluxes_h001}). Most of the extracted energy flows into
the disc, and very little in the polar region. 

The fourth panel shows the (negative) binding energy brought inward with
the gas. This effectively transports energy outward. This component is
more isotropic than the net energy flux, reflecting the fact that gas
falling in along the axis is more bound than gas accreting in the equatorial
plane.

The fifth panel shows the magnetic component, which, as
we argued in the previous section, corresponds roughly to the viscous
energy transfer rate. As in the case of a thin disc, viscosity
transports energy outward and redistributes it throughout the
disc. Most of this energy goes near the equatorial plane, where the
density is largest, and viscosity most efficient.

Finally, the bottom-most panel presents the flux of internal energy. Its
magnitude is significant (compare
Fig.~\ref{f.enfluxes_h001}) because optically thin flow cannot cool
efficently and becomes very hot. As expected, thermal energy is
brought inward with the gas, and the angular distribution is again
quasi-spherical -- although the accretion rate is highest near the
equatorial plene, the gas temperatures there are lower than in the
polar region.

\subsection{Energy flow in optically thick super-critical discs}
\label{s.slim}

Accretion flows transferring gas at rates higher than the Eddington
limit are optically thick, but they are not as radiativelly efficient
as thin discs. The vertical optical depth is so large, that the
cooling time becomes comparable or larger than the accretion time, and
such discs cannot cool efficiently. Instead, significant fraction of
radiation may be advected on to a black hole. We now look closely at
the simulation \texttt{r001} of a mildly super-critical disc
($10\medd$) near a non-rotating BH.

\subsubsection{Luminosities}

\begin{figure}
\includegraphics[width=.95\columnwidth]{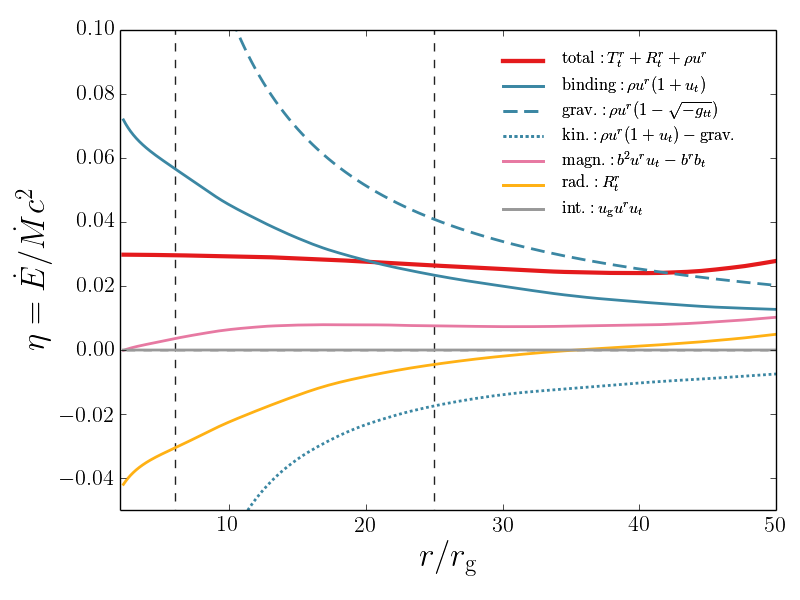}
\caption{Total luminosity and its components for a GR radiative MHD
  simulation of optically thick, super-critical disc accreting at
  $\sim 10\medd$ on a non-rotating BH (model \texttt{r001}). The
  colors have the same meaning as in Fig.~\ref{f.enfluxes_h001}.}
\label{f.enfluxes_r001}
\end{figure}

Figure~\ref{f.enfluxes_r001} shows the luminosities in various
components of energy for the super-critical disc \texttt{r001}. It
can be directly compared to Fig.~\ref{f.enfluxes_h001} corresponding
to an optically thin disc. The two figures show qualitatively the same
behavior of the binding and magnetic components  -- bound gas is
brought inward and effectively transports energy outward, similarly to the
magnetic/viscous energy flux which takes energy liberated in the innermost
region and redistributes it in the outer regions. 

There is, however, a
significant difference in the thermal and radiative components. In the
case of an optically thin disc, the liberated energy heats up the
gas and results in significant inward flux of thermal energy (grey
line in Fig.~\ref{f.enfluxes_h001}). In the case of a radiative
super-critical disc, the flux of thermal energy is
negligible. Instead, the radiative component is now significant. It is
negative within $r\approx35r_g$, reflecting the fact that photons are
trapped in the optically thick flow 
and transported with the gas to the BH. Only outside this radius, the amount of
radiative energy flowing out exceeds the advected one. The fact that
the thermal energy is now subdominant with respect to the radiative
energy is consistent with the fact that super-critical discs are
radiation-pressure dominated, and therefore, inward motion advectively
carries radiative energy, not thermal energy.

Interestingly, the total luminosity of the optically thick disc is
again close to $3\%\dot M c^2$ (thick red line in
Fig.~\ref{f.enfluxes_r001}). This is the amount of energy returned
from the system to the ISM in the case of super-critical accretion on
to a non-rotating black hole. Noticeably, the magnitude of BH
feedback power from a geometrically thick disc near a non-rotating BH is not sensitive to its
optical depth.

\subsubsection{Angular distribution}

\begin{figure}
  \rotatebox{90}{\hspace{1.2cm}Density}\hspace{.05cm}\includegraphics[width=1.0\columnwidth]{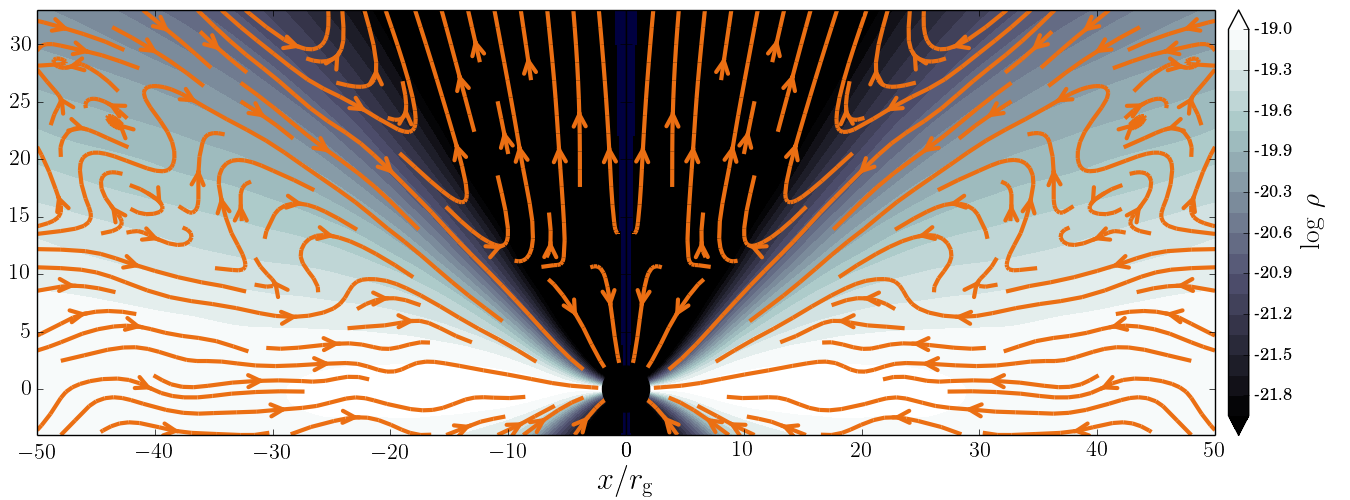}\vspace{-.3cm}
  \rotatebox{90}{\hspace{1.3cm}Rest mass}\hspace{.05cm}\includegraphics[width=1.0\columnwidth]{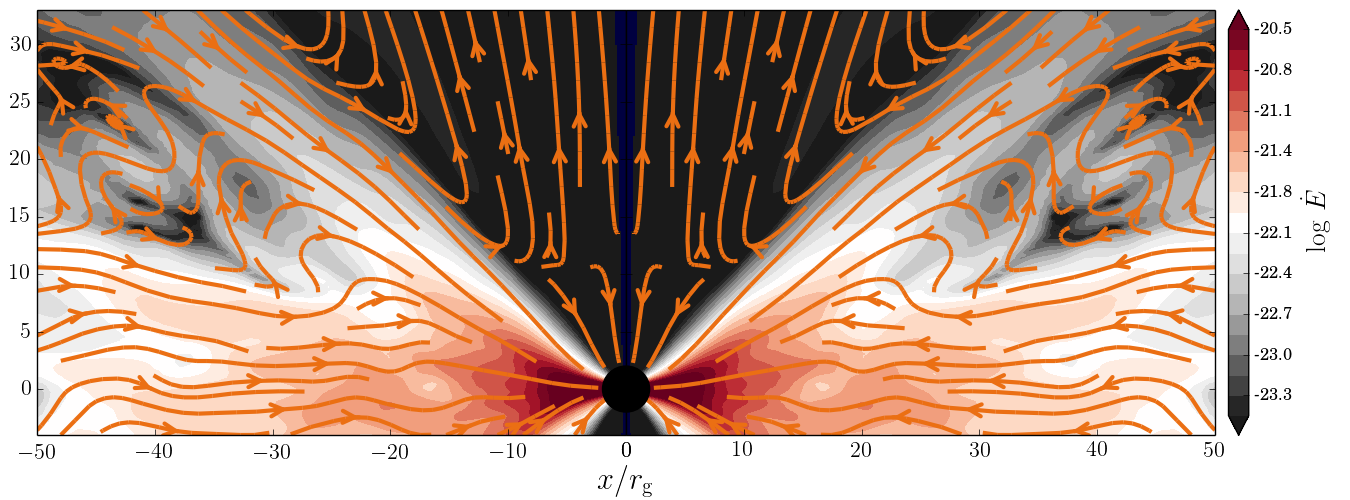}\vspace{-.3cm}
  \rotatebox{90}{\hspace{1.4cm}Total}\hspace{.05cm}\includegraphics[width=1.0\columnwidth]{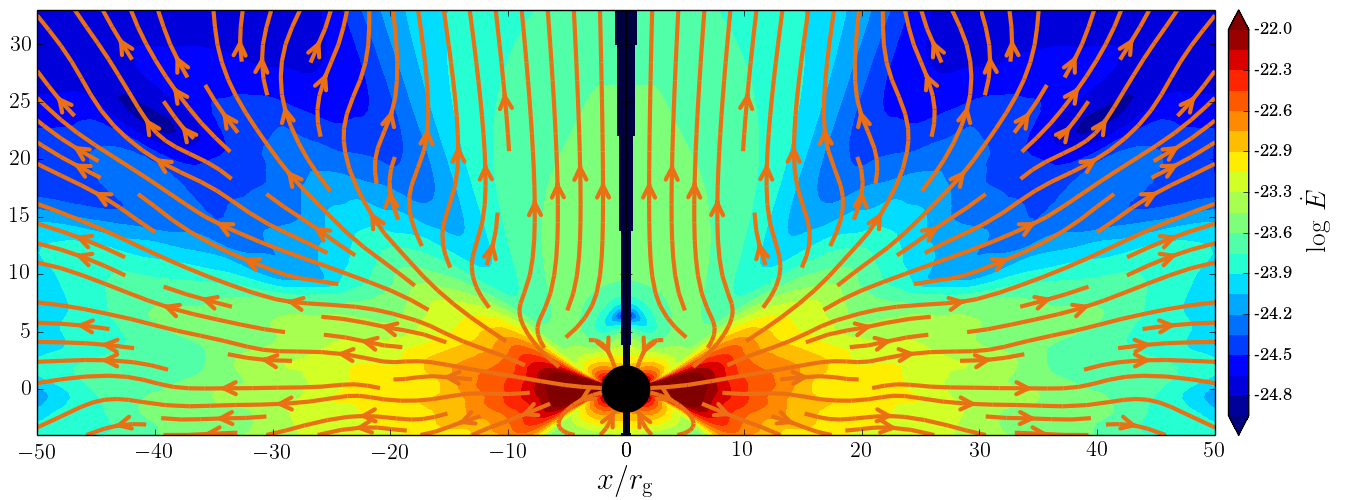}\vspace{-.3cm}
\rotatebox{90}{\hspace{1.3cm}Binding}\hspace{.05cm}\includegraphics[width=1.0\columnwidth]{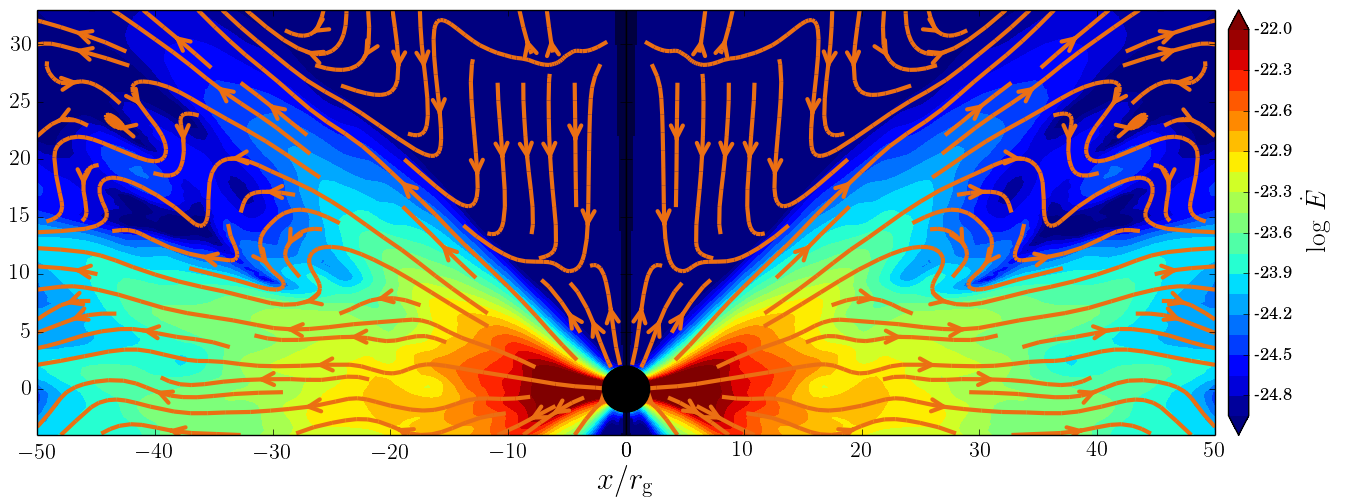}\vspace{-.3cm}
\rotatebox{90}{\hspace{1.cm}Magnetic/Visc.}\hspace{.05cm}\includegraphics[width=1.0\columnwidth]{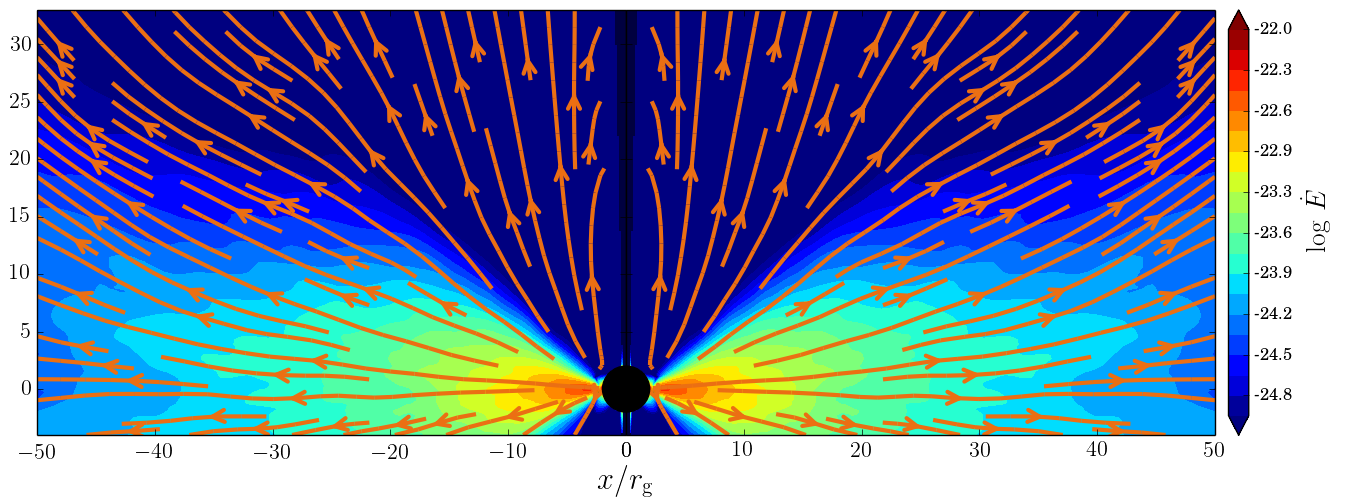}\vspace{-.3cm}
\rotatebox{90}{\hspace{1.25cm}Radiative}\hspace{.05cm}\includegraphics[width=1.0\columnwidth]{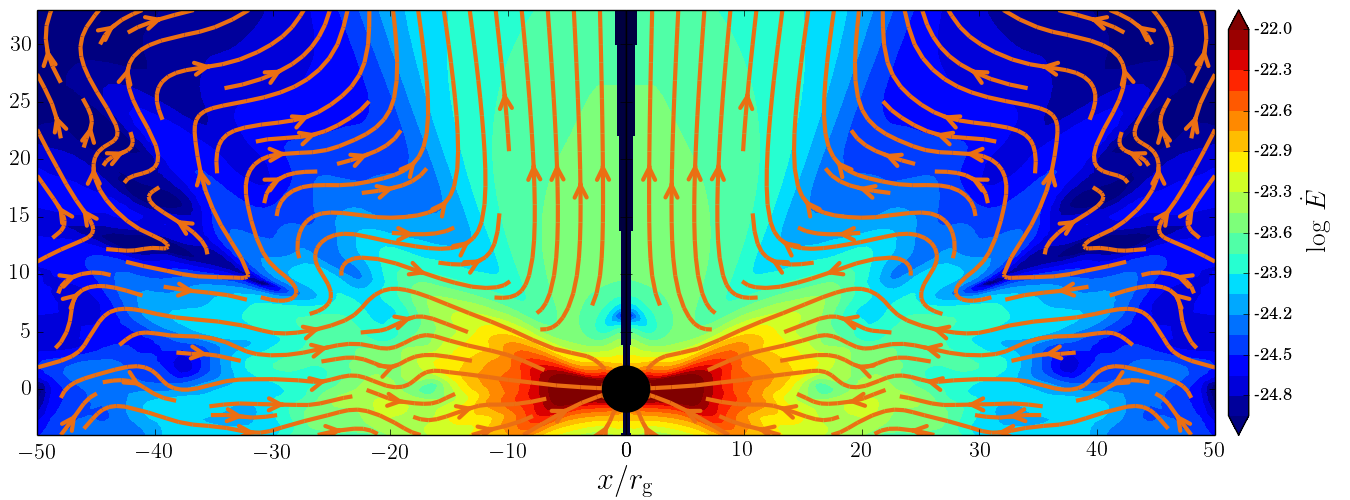}
\caption{Similar to Fig.~\ref{f.sims_h001} but for a super-critical,
  optically thick disc (model \texttt{r001}). The bottom-most panel
  now shows the flux of energy carried by radiation. The flux of
  thermal energy is negligible.}
\label{f.sims_r001}
\end{figure}

The distribution of the energy fluxes on the poloidal plane for the
optically thick simulation \texttt{r001} is shown in
Fig.~\ref{f.sims_r001}. The density distribution (top-most panel)
shows much larger contrast between the equatorial plane and the polar
axis than in the case of an optically thin disc. This fact results
from radiative pressure exerted on the gas in the funnel --
gas is accelerated vertically and escapes along the axis. This is
clearly reflected in the velocity streamlines shown in that panel.

The total energy extracted from the system (third panel) looks
different than in the previous case. This time the polar region is not
empty of outflowing energy. The optically thin radiation escaping
along the polar funnel dominates the energy budget
there.

The distributions of binding and magnetic energy fluxes are similar to
the optically thin case. Both transfer energy within the bulk of the
disc. For the case of the magnetic energy flux (which reflects the
effective viscous transport), this fact supports the conjecture that
this energy will dissipate at larger radii (it would not dissipate if
the magnetic energy has left the disc, e.g., along the axis).

Finally, the bottom-most panel shows the magnitude and direction of
the radiative flux. As already mentioned, radiation manages to escape
in the polar region. However, it is trapped in the optically thick
flow near the equatorial plane.

\subsection{Higher accretion rates}
\label{s.highmdot}

Here, we briefly discuss how the picture described in the previous
Section changes when the accretion rate increases but the BH spin remains zero. Detailed
comparison of models \texttt{r001} (accreting at $10\medd$) and
\texttt{r003} ($176\medd$) was given in \cite{sadowski+3d}. The most
important points are as follows.

\begin{figure}
\includegraphics[width=.95\columnwidth]{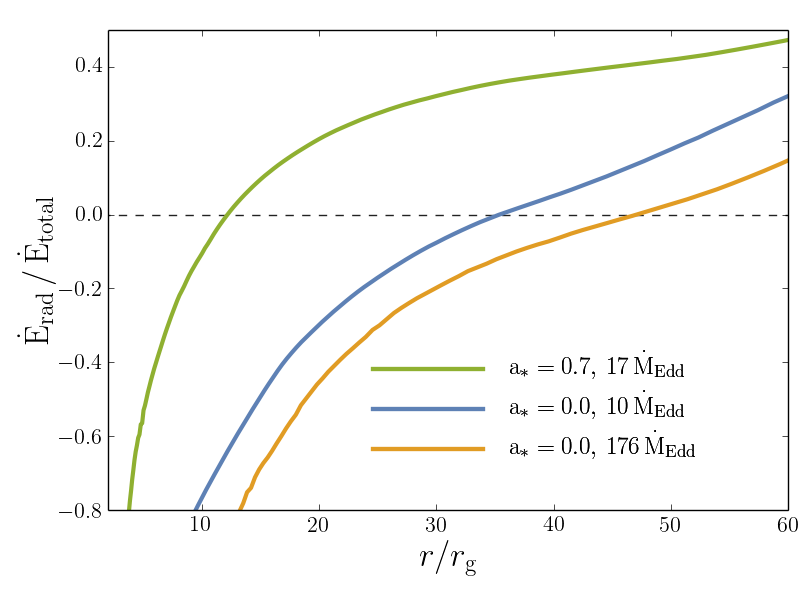}
\caption{Fractional contribution of the radiative luminosity to the
  total luminosity in simulations \texttt{r011} (green), \texttt{r001}
(blue), \texttt{r003} (orange lines). The luminosities were obtained
by integrating corresponding fluxes over the whole sphere. The
radiative luminosity, in particular, includes both the radiation
trapped in the flow and escaping to infinity.}
\label{f.radflux_all}
\end{figure}

The total efficiency in both cases equals approximately
$3\%$. However, because of the larger optical depth in model
\texttt{r003}, photon trapping is more effective. In particular, the
polar region becomes optically thick. Inside $r\approx 30$ the gas is
dragged on to the BH even along the axis. As a
result, radiative luminosity of the system goes
down. Fig.~\ref{f.radflux_all} shows the fractional contribution of
the radiative luminosity, $L_{\rm rad}$, to the total luminosity
$L_{\rm tot}$. Blue and orange lines correspond to simulations
\texttt{r001} and \texttt{r003}, respectively. It is clear that the
latter is less radiatively luminous, and that the effective trapping
radius moves outward. This fact, however, turns out not to change the
total efficiency.

\subsection{Rotating black hole}
\label{s.rotating}

So far we have been discussing accretion flows around non-rotating BHs. In
this Section we briefly discuss what impact  non-zero BH spin has on the
energy flow properties.

BH spin affects accretion flows in two ways. Firstly, BH rotation
modifies the spacetime geometry and for a given BH mass allows for circular 
orbits getting closer to the horizon with increasing BH spin. This fact results in an increased efficiency of accretion
-- the closer is the inner edge of the disc, the more binding energy
is liberated. Secondly, BH rotational energy can be extracted in the
Blandford-Znajek process \citep{bz}. The power of the related jet
 depends on the value of the BH spin and on the amount of the magnetic
 flux that has been accumulated at the horizon. The latter is
 determined by the geometry of the magnetic field in the accreted gas and
 the efficiency of magnetic field dragging. It is known
 not to exceed the value characteristic for the magnetically arrested
 (MAD) state \citep{igu+03,narayan+mad,tch+11}.

In this paper we analyze one simulation (\texttt{r011}) of super-critical accretion
on a mildly-rotating (the non-dimensional spin parameter $a_*=0.7$)
BH. The average accretion rate in this run ($17\medd$) is comparable
to the fiducial simulation \texttt{r001}, and allows for direct
comparison. The amount of the magnetic flux accumulated at the BH (the
magnetic flux parameter $\Phi \approx 15$) is
far from the MAD limit of $\Phi \approx 50$, but is large enough to study the impact of
the extracted rotational energy. Optically thick super-critical accretion
flows in the MAD limit were studied in a recent work by
\cite{mckinney+madrad}.

The total efficiency of simulation \texttt{r011} is roughly $8\%$,
significantly higher than that of the comparable simulation on a
non-rotating BH ($3\%$). The increase in efficiency comes from both
factors mentioned above (modified spacetime geometry and the
extraction of the rotational energy). The latter by itself should
extract $\sim 6\%\dot M c^2$ for the accumulated amount of magnetic
flux and spin $a_*=0.7$, but decomposition of the total energy into
these two components is not straightforward.

\begin{figure}
  \rotatebox{90}{\hspace{1.2cm}Density}\hspace{.05cm}\includegraphics[width=1.0\columnwidth]{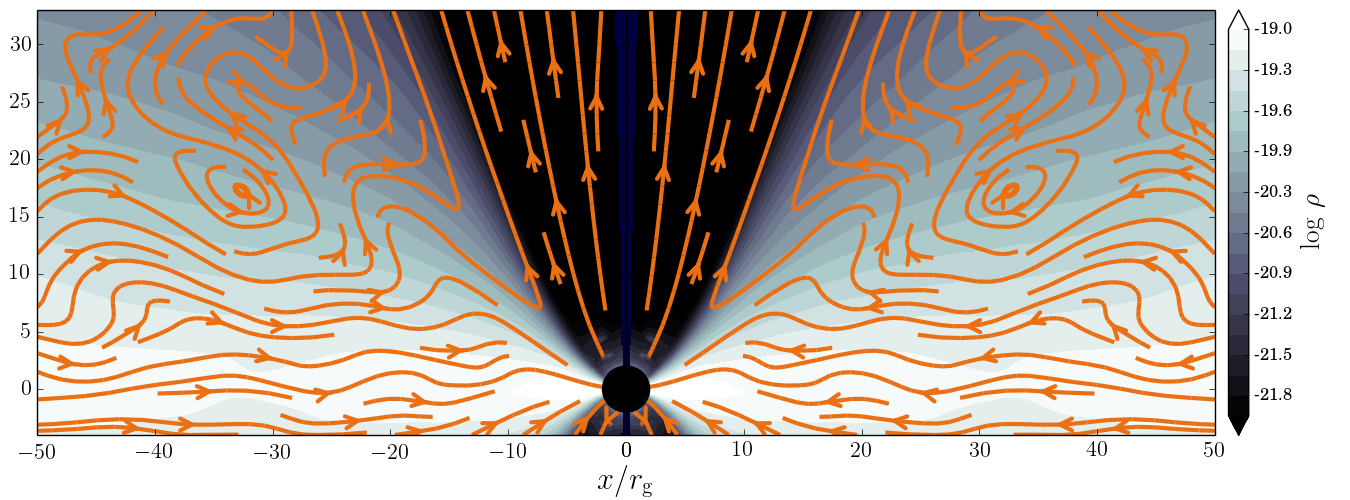}\vspace{-.3cm}
  \rotatebox{90}{\hspace{1.3cm}Rest mass}\hspace{.05cm}\includegraphics[width=1.0\columnwidth]{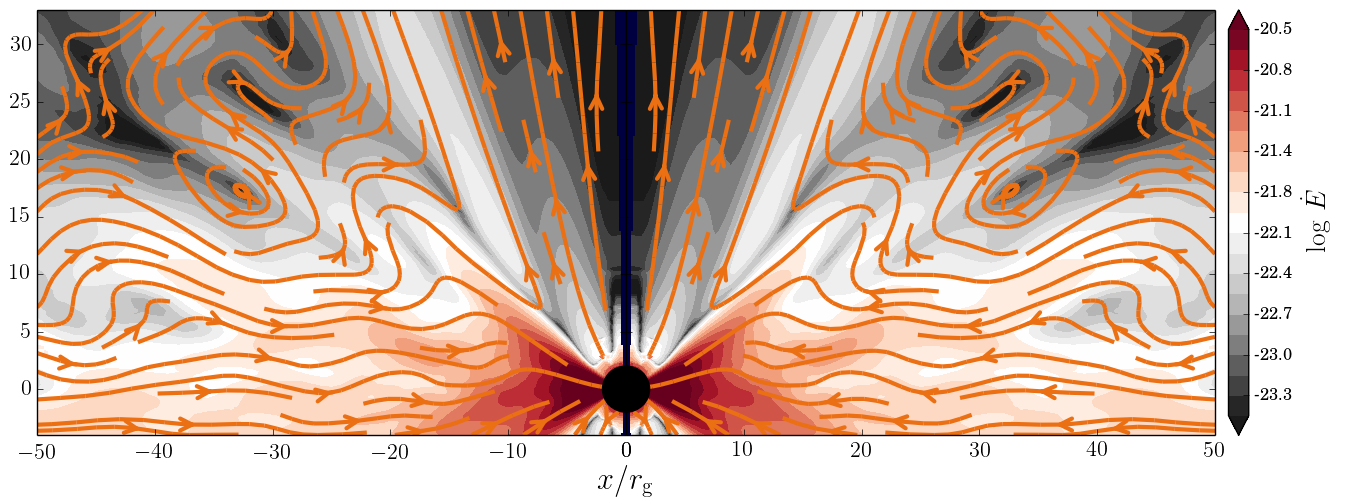}\vspace{-.3cm}
  \rotatebox{90}{\hspace{1.4cm}Total}\hspace{.05cm}\includegraphics[width=1.0\columnwidth]{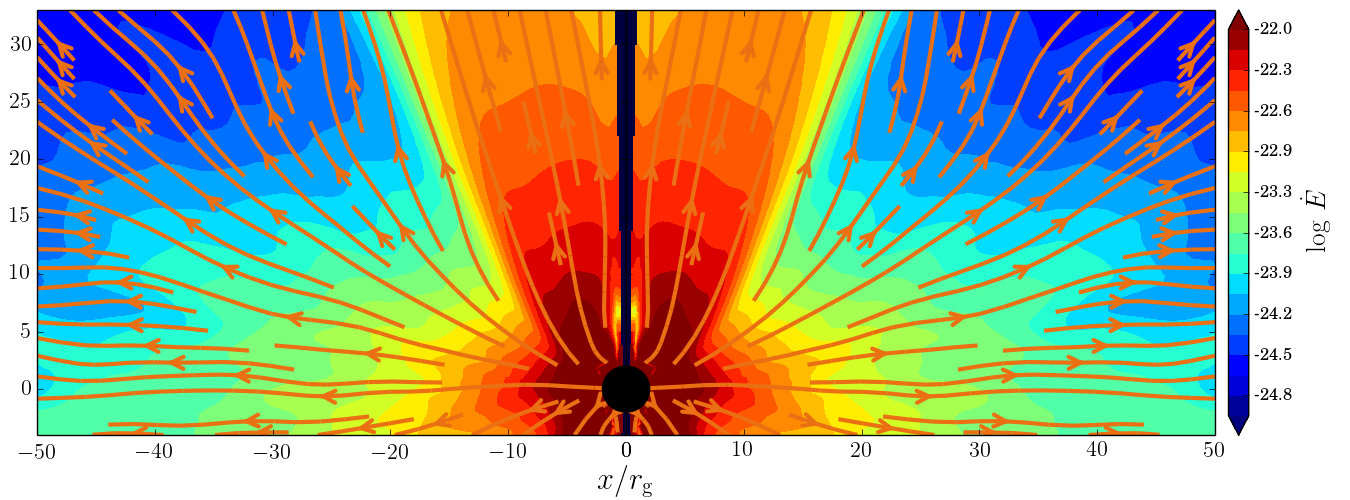}\vspace{-.3cm}
\rotatebox{90}{\hspace{1.3cm}Binding}\hspace{.05cm}\includegraphics[width=1.0\columnwidth]{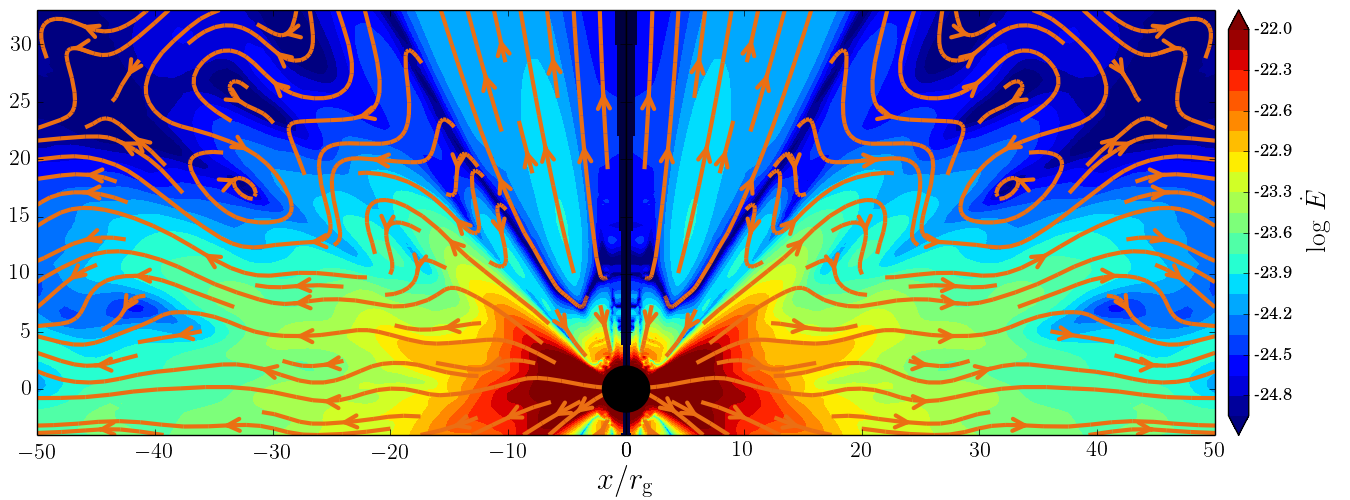}\vspace{-.3cm}
\rotatebox{90}{\hspace{1.cm}Magnetic/Visc.}\hspace{.05cm}\includegraphics[width=1.0\columnwidth]{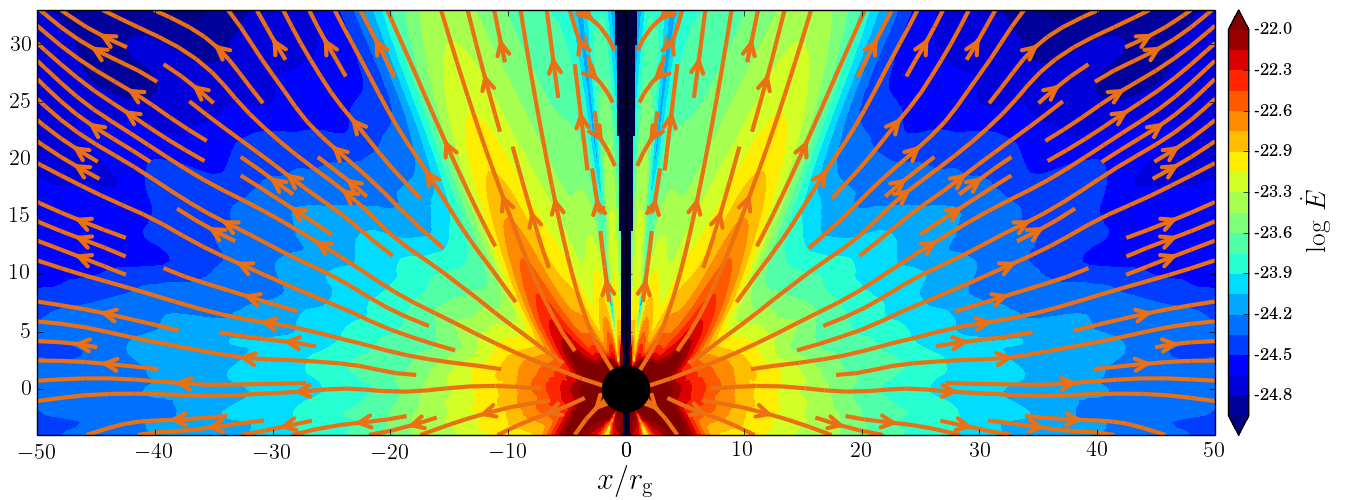}\vspace{-.3cm}
\rotatebox{90}{\hspace{1.25cm}Radiative}\hspace{.05cm}\includegraphics[width=1.0\columnwidth]{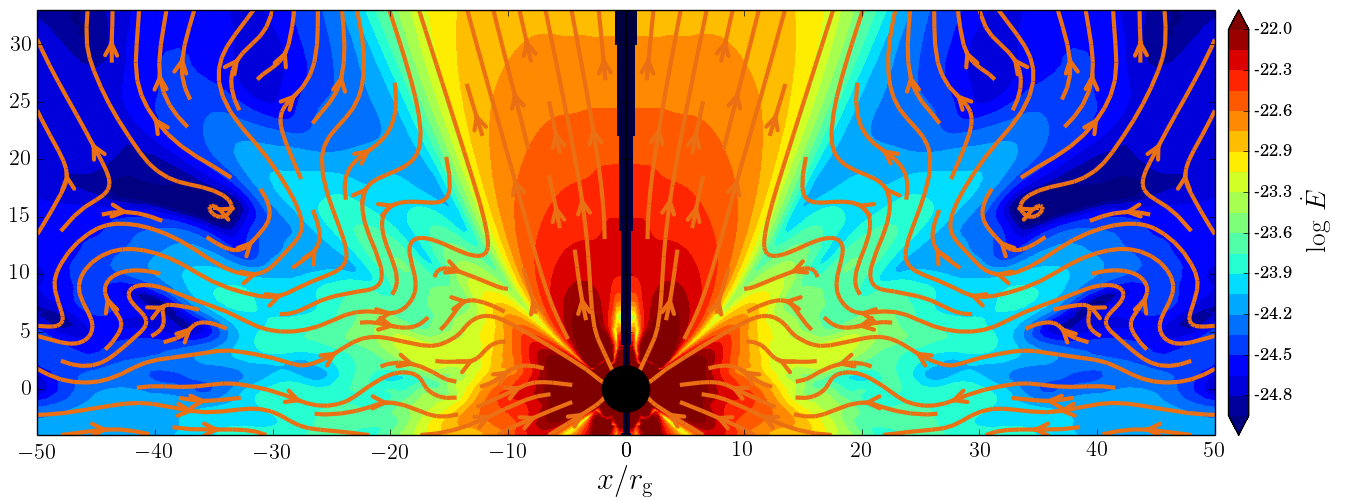}
\caption{Similar to Fig.~\ref{f.sims_r001} but for a model with
  spinning BH (model \texttt{r011}).}
\label{f.sims_r011}
\end{figure}

\begin{figure}
\rotatebox{90}{\hspace{1.25cm}Radiative}\hspace{.05cm}\includegraphics[width=1.0\columnwidth]{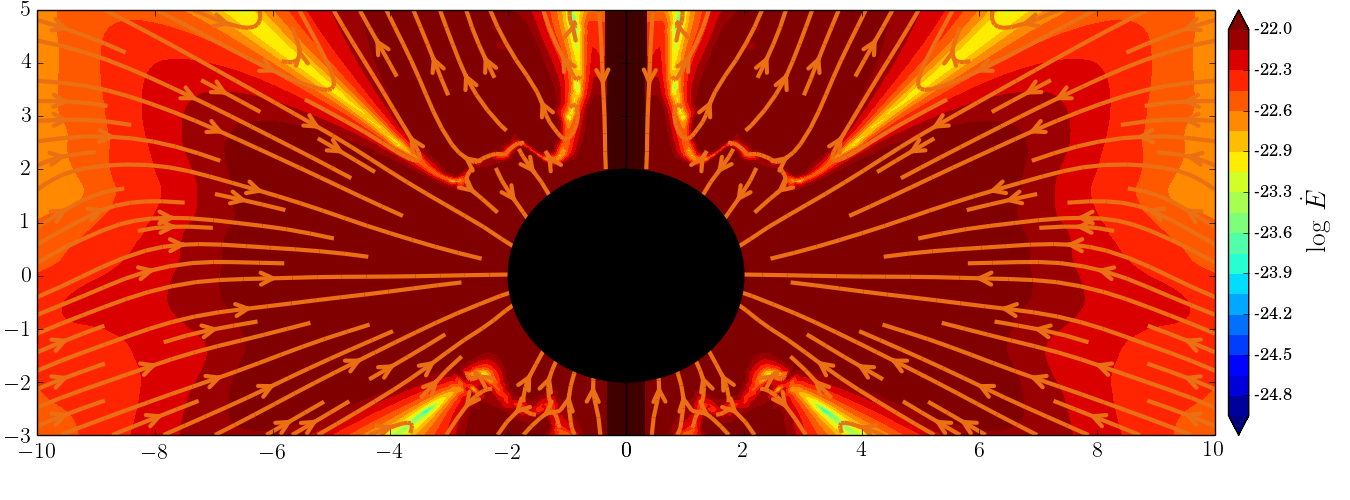}
\caption{Zoomed in magnitude and direction of radiative energy flux for a model with
  spinning BH (model \texttt{r011}).}
\label{f.sims_zoomin_r011}
\end{figure}

In Fig.~\ref{f.sims_r011} we show the distribution of energy flux
components on the poloidal plane. The panels have the same meaning as
in the previously discussed Fig.~\ref{f.sims_r001}. There are a couple
of  noticeable  differences between the two. Most importantly, the amount
of total energy extracted into the funnel region is much higher for
the rotating BH case. This is expected, because the energy extracted
in the Blandford-Znajek is known to go roughly along the axis
\citep{penna+membrane}. In the case of optically thin accretion on to a
rotating BH \citep[e.g.,][]{tch+12,sadowski+outflows}, the jet power is
extracted as magnetic Poynting flux gradually converting (if mass
loading is significant) into kinetic energy of gas. In the case of the
radiative flow studied here, this extra energy is carried mostly by
radiation already for $r\gtrsim 3r_g$. Magnetic flux is significant only
in a shell surrounding the funnel region. Fig.~\ref{f.sims_zoomin_r011} shows the magnitude and
direction of the radiative flux in the immediate vicinity of the
BH. As expected, radiative flux falls \textit{on the BH} in this innermost
region, and it is the magnetic
energy which is extracted at the
horizon. However, the latter is quickly converted into the radiative
energy. This is possible because the magnetic field efficiently pushes
hot and optically
thick gas along the axis. The gas, in turn, drags the radiation
upward.

At the risk of oversimplifying, it is possible to say that the
properties of an energy flow in the case of a super-critical accretion on to a
rotating BH are a superposition of the disk component (discussed in Section~\ref{s.slim} for  a non-rotating BH)
and the jet contribution coming from the Blandford-Znajek
process. The power of the latter depends on the BH spin and magnetic
flux threading the horizon, and may overwhelm the former in magnitude.
At the same time, the jet component is limited only to the polar
region. If the confinement provided by the
disc is strong enough, it is likely to stay collimated. 

\section{Discussion}
\label{s.discussion}

\subsection{The fate of the energy flow}
\label{s.fate}

We have shown so far that in geometrically thick discs, both optically
thin and thick, a significant
flux of energy is liberated in the accretion flow and flows out of the
system. Although the simulations we performed allowed us to study only the
innermost ($r\lesssim 25$ at the equatorial plane) region of the
flow\footnote{Obtaining an equilibrium solution in a larger domain is
  computationally very demanding because of the increasing range of
  timescales. An appropriate approach would be to divide the domain
  into subregions which are simulated independently but coupled at the
  boundaries \citep[e.g.][]{yuan+12}.},
we were able to infer the total luminosity of the system. This is because in a stationary state this
quantity is determined by the energy flux 
crossing the BH horizon.

However, the adopted method does not allow to study what happens to
the extracted energy outside the inflow/outflow equilibrium region
(i.e., for $r\gtrsim 25$) -- only for gas inside this region the
duration of the simulations was longer than the viscous time scale.

As Figs.~\ref{f.enfluxes_h001} and \ref{f.enfluxes_r001} show, the
total luminosity of the flows around a non-rotating BH (roughly $3\%\dot
Mc^2$) comprises three components at radius
$r=25$. The largest in the magnitude is
the binding energy flux which effectively
deposits energy at infinity. The magnetic component (reflecting 
the viscous energy transport) is also transporting energy outward in a
significant amount. The remainder goes either into the thermal (for optically thin case) or radiative energy (for
optically thick) component. In both cases their net effect results in advecting energy
inward.

The binding energy consists of the gravitational and kinetic
components (plotted with dashed and dotted blue lines, respectively, in
Figs.~\ref{f.enfluxes_h001} and \ref{f.enfluxes_r001}). The former
goes to zero with increasing radius, and \textit{at infinity} no
gravitational energy is transported by the gas. The
kinetic component is negative inside the computational domain
reflecting the fact that gas flows inward and carries kinetic energy
of its rotational motion. However, when outflows are efficiently generated it might become ultimately positive outside the computational domain.

As we have shown above,
the radial flux of magnetic energy reflects the effective 
viscous energy transfer. Viscosity not only
transports angular momentum and energy, but also leads to dissipation
of the latter. Therefore, one may expect that the amount of energy
carried by magnetic fields will dissipate sooner or later
outside the convergence region of the simulation, adding up to the
local heating rate in the same way as for thin discs discussed in
Section~\ref{s.thin}. Thus, no magnetic energy will be ultimately directly
deposited in the ISM.

The radiative energy transfer is important only for optically thick
accretion flows. Simulations of such flows described in this paper (models
\texttt{r001} and \texttt{r003}) show
significant photon trapping in the bulk of the disc which results in
negative net flow of radiative energy (see Figs.~\ref{f.enfluxes_r001}
and \ref{f.radflux_all}) in the inner region. However, as radiation
gradually diffuses out from the disc, the outflowing component finally
overcomes inward advection, and the net radiative luminosity
becomes positive. In particular, from the point of view of an observer
at infinity, radiation will only carry energy outward. In the
following Section we discuss how bright the accretion flow can be.

The thermal energy flux, which contributes
significantly to the energy transfer rate in optically thin accretion
flows, reflects both the advective and convective contributions. In
the inner part of the flow it is the
advective component which dominates and results in negative net energy
transfer rate -- hot gas is accreted and takes its thermal
energy with it. However, if outflows are present in the outer region,
the net effect may be opposite and the thermal energy 
carried advectively outward with the outflowing gas may dominate.

Convection can carry energy against the gravity without
transporting mass. This component is negligible in the simulations we
performed (thermal energy flows advectively inward), but in principle it may
become significant, or even dominant, in the outer region. Several
models of accretion flows which are dominated by convection
(convectively dominated accretion flows, CDAFs) have been formulated
\citep{quataert-cdafs,narayan-cdafs,narayan-cdafs2,abramowicz-cdafs}. Whether convection is
important is an open question. \cite{narayan+12} performed a set
of simulations similar to our model \texttt{h001} and showed that
optically thin flows are convectively stable within $r\lesssim
100$. On the other hand, \cite{yuan+15}, who studied
    non-magnetized viscous flows on large
    scales, found that the
    accretion flow is actually convectively
    unstable. Even if convection is important, it cannot transport
    energy beyond the
    outer edge of the disc (or beyond the Bondi radius). One may expect that ultimately all energy
    transported by convection is released as radiation or generates outflow.

  To sum up, geometrically thick accretion flows on to a non-rotating
  BHs deposit in the ISM roughly $3\%$ of the rest-mass energy
  crossing the BH horizon. This energy may be transported outward with
  the outflowing gas, radiation and convection. Which components
  dominate is currently unclear. Ultimately, however, only radiation
  and outflow can transport energy beyond the Bondi radius and 
  deposit it in the ISM. A separate jet component may be
  present in case of a spinning BH which managed to accumulate
  significant magnetic flux at its horizon. It will result in a
  collimated, narrow outflow of mostly kinetic energy, unlikely to
  interact efficiently with the ISM.

\subsection{Radiative luminosity}
\label{s.luminosity}

Radiation is one of the ways of extracting energy from accretion
flows. The total luminosity (accounting for all forms of energy) for a thick disc near a non-rotating
BH seems to be robust --
every simulation we have performed indicates that roughly $3\%$
of the accreted rest-mass energy is returned to ISM. The amount of
this energy that goes into radiation is not, however, easy to
estimate. Only when the photosphere is properly resolved, one can
check  if radiation reaches the observer at infinity. This  is not
the case for any of the optically thick simulations discussed here. They were
run only for a time which allowed them to reach inflow/outflow
equilibrium state at the equatorial plane within $r\approx
25$. Because of large optical depths, photospheres are located at
large distances \citep{sadowski+dynamo}, significantly outside the
converged region. This fact makes it impossible to directly measure
the amount of radiation escaping the system. Only radiation
escaping along optically thin funnel, if  it  exists, is guaranteed to
reach a distant observer.

Because of significant photon trapping in the
super-Eddington regime, the radiative luminosity of
the system is not proportional to the accretion rate. What is more,
the radiation coming from such an accretion flow must penetrate
the optically  thick wind region. It cannot be therefore locally significantly
super-Eddington, because in such a case it would transfer its energy
and momentum to the gas accelerating it. We are inclined to suggest that
the result will be similar to the effect of pure photon trapping
which results in logarithmic dependence
of the luminosity on the accretion rate (already anticipated by
\cite{ss73}, see also \cite{begelman-79}),
\be
L_{\rm rad}\approx L_{\rm Edd}\left(1+\log \dot M/\medd\right).
\label{e.Lradestimate}
\ee

A similar logarithmic behavior was found in the old
works on super-Eddington Polish Doughnuts and
explained by Paczynski and collaborators as being
a consequence of the drop in efficiency
when, with increasing accretion rate, the inner edge of the
accretion disc moves from the ISCO to the innermost bound circular orbit (IBCO), where the efficiency
is zero \citep[see][for explanation and
references]{wielgus+15}.

\subsection{Outflow}

Our work shows that the existence of outflows is inevitable in outer parts of thick
accretion discs. Avoiding them requires \textit{all} of the extracted
energy to be ultimately
transported outwards by radiation. However, thick discs
are radiatielly inefficient (see
Eq.~\ref{e.Lradestimate}).

This conclusion is based on disc energetics --
a significant fraction of the accreted rest mass energy flows outward
through the disc which cannot generate
enough radiation to provide the efficient cooling required
to get rid of the energy surplus. If convection is not effective, at
least in the outermost region, outflow is the only possible way of taking
this excess of energy out of the system.

We do not see strong outflows in the simulations we
performed (compare
topmost panels in Figs.~\ref{f.sims_h001}, \ref{f.sims_r001}, and
\ref{f.sims_r011}). Only in the funnel
region of optically thick simulations \texttt{r003} and \texttt{r011} one observes
that the radiative luminosity is converted gradually into kinetic
energy of the outflowing gas. However, the kinetic luminosity of such gas measured at the
outer boundary is still at most $\sim 10\%$ of the total 
efficiency.

Therefore, one may expect that most of the outflow will be generated
at radii larger than covered by the inflow/outflow equilibrium region
of the simulations, i.e., at $r\gtrsim 25r_g$. What will drive these
outflows? In principle, there are three acceleration mechanisms likely
to act in magnetized accretion flows -- magnetocentrifugal
\citep{blandfordpayne}, radiative and thermal. Magnetocenrifugal
driving is not effective in the simulated inner region of a non-MAD
accretion flow \citep[see
also][]{moller+15}, and there is little hope for it to
become effective further out, where magnetic field has no reason to
be more uniform on large scales. Radiative driving is seen in the
funnel region of the simulated super-critical discs, but does not
result in 
significant outflow
at larger polar angles -- radiation diffusing out of the disc into the
optically thick wind region is supporting the disc against gravity,
and therefore cannot on average significantly exceed the local Eddington flux. 

Thermal wind driving remains the only candidate to balance the energy
budget of the discs. It is especially reasonable if we consider the
energy flux redistributed through viscosity. When it finally
dissipates at larger radii, it will heat up the gas and make it more
prone to become unbound and likely to flow out of the system. This is
in agreement with the standard ADAF model which predicts positive
Bernoulli function for the inflowing gas in the self-similar regime \citep{narayanyi-94}. At
the same time, most of the observed outflows in BH accretion flows are
believed to be of thermal nature \citep[e.g.,][]{lee+02,ponti+12,neilsen-13}.

The total
  luminosity will be ultimately carried by the outflow and
  radiation. Thus, at infinity, the outflow will carry the amount of
  energy equal to the difference between the total and radiative
  luminosities. For a non-spinning BH one will have, \be L_{\rm outflow,\infty}= 0.03\dot Mc^2-L_{\rm
    rad,\infty}.  \ee The latter term is obviously negligible for
  optically thin discs, for which the whole extracted energy goes into
  outflow.

\subsection{Extent of a thick disc}

In the considerations so far we assumed that the accretion flow
extends to infinity, and that the gas at infinity has negligible
energy, i.e., its Bernoulli number is zero. These conditions do not
have to be satisfied in reality. In particular, a thick disc is
expected to become thin in the outer region. For example, a
super-critical disc becomes radiatively efficient at a sufficiently
large radius where its optical depth is no longer large enough to prevent
locally generated radiation from diffusing out in an efficient
way. Somewhat similarly, optically thin discs (ADAFs) cannot exist
above a critical accretion rate, and this limit decreases with
radius. Therefore, for a fixed accretion rate at a BH, there is a
radius where thick disc must become thin, and radiatively
efficient. In such cases, the picture presented here would have to be
modified accordingly, e.g., if at the transition radius the outward
energy flux inside the disc (i.e., not in the outflow) is still significant, then one could
expect that the thin disc, extending from this radius outward, will
ultimately release all its energy as (relatively cold) radiation.

\subsection{Transition between accretion modes}

We did not simulate thin, sub-Eddington
accretion flows, for which the standard assumption of being radiatively
efficient is satisfied by construction. For such discs, the energy
transfer is expected to follow the characteristics described in
Section~\ref{s.thin}, i.e., the total efficiency of BH feedback equals
the thin disc efficiency, and ultimately all of it is carried by
radiation which is emitted over a wide solid angle. Geometrically thin discs
are unlikely to drag significant amount of magnetic field on to the BH
(e.g., \cite{lubow+94}, \cite{ghosh+97}, \cite{guletogilvie-12,guletogilvie-13}, but see also
\cite{spruit+05}, \cite{roth+08} and \cite{beckwith+09})
and therefore one does not expect strong Blandford-Znajek jet
component.

The transition between optically thick but geometrically thin and
thick discs takes place near the Eddington accretion rate. In the past
it has been modeled with the so called slim disc model
\citep[e.g.,][]{abramowicz+88,sadowski.phd} which
generalizes the standard thin disc model to higher accretion
rates. Recently, numerical simulations similar to the ones that this
work is based on, have studied a number of super-Eddington accretion
flows \citep[e.g.,][]{sadowski+dynamo,jiang+14b}. Simulations of thin
discs, more demanding computationally, have not yet been
performed. Below, we
will describe the transition between geometrically thin and thick optically thick
discs with the help of arbitrary step functions, which make the
final formulae agree qualitatively with what we have learned from
numerical, multi-dimensional simulations.

The transition between optically thin and thick discs is even less well
understood and awaits numerical modeling. It is known than
radiatively inefficient optically thin flows cannot exist above some
critical accretion rate $\dot M_{\rm ADAF}\approx 10^{-3}\Medd$ \citep[e.g.,][]{esin+97}. Whether
increasing the accretion rate above this threshold results in a dramatic
transition to a cold, optically thick disc, or rather the disc
takes a form similar to the luminous-hot accretion flow
\citep[LHAF,][]{yuan+01}, has still to be verified. 

Below,
for simplicity, we assume that whenever accretion rate is below $\dot
M_{\rm ADAF}$, accretion occurs in optically thin disc, and that the
transition to optically thick discs (for  $\dot
 M>\dot M_{\rm ADAF}$) takes place instantanously.

Having these considerations in mind, one may
approximate the total amount
of feedback luminosity coming from an accreting system as,
\bea
\hspace{1cm}L_{\rm fb}=\frac12\eta_{\rm thin}\dot Mc^2 + P_{\rm BZ},
\label{e.Lfbthin}
\eea
\hspace{4cm}for $\dot M<\dot M_{\rm ADAF}$ (opt. thin),
\bea
\hspace{1cm}L_{\rm fb}=\eta_{\rm thin}\left(1-\frac12f_\eta\right) \dot
  Mc^2+f_{\rm BZ} P_{\rm BZ},\label{e.Lfb3}
\label{e.Lfbthick}
\eea
\hspace{4cm}for $\dot M>\dot M_{\rm ADAF}$ (opt. thick),
\vspace{.3cm}

\noindent where $\eta_{\rm thin}$ stands for the efficiency of a standard thin
disc with given spin, the $1/2$ factor reflects two times smaller
efficiency of thick discs, and where we allow for the Blandford-Znajek
contribution, $P_{\rm BZ}$, for thick discs. $\dot M_{\rm ADAF}\approx
10^{-3}\medd$ is
the critical accretion rate above which radiatively inefficient optically thin accretion flows do
not exist. Functions
$f_\eta$ and $f_{\rm BZ}$,
\be
f_\eta=\left(1+\left(\frac{3}{\dot
    M/\medd}\right)^3\right)^{-1}
\label{eq.feta}
\ee
\be
f_{\rm BZ}=\left(1+\left(\frac{1}{\dot
    M/\medd}\right)^5\right)^{-1}
\label{eq.fbz}
\ee
were chosen to give (arbitrary) smooth transitions
between the sub- and super-Eddington regime for the efficiency and the
jet power, respectively. The Blandford-Znajek
term \citep[given here for
saturated magnetic field at the BH, i.e., for the MAD limit, see][]{tchekh15},
\be
 P_{\rm BZ}=1.3a_*^2\dot M c^2,
\label{e.PBZ}
\ee is strongly damped for thin discs which are not likely
to drag the magnetic fields effectively.

\begin{figure}
\includegraphics[width=.985\columnwidth]{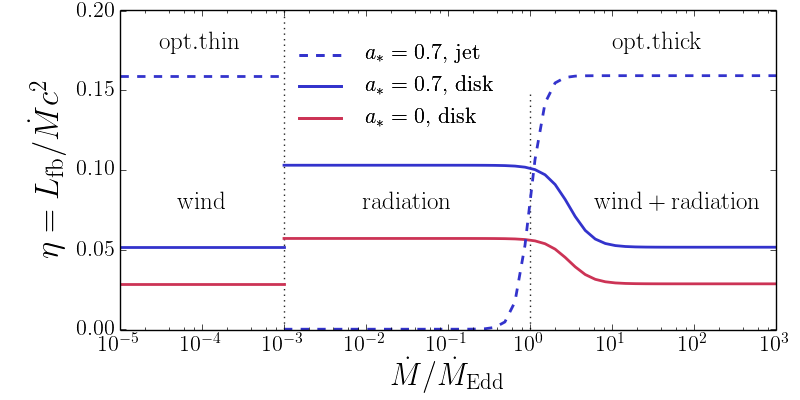}
\caption{Total efficiency of the feedback (Eqs.~\ref{e.Lfbthin} \&
  \ref{e.Lfbthick}) for $a_*=0.0$ (red) and $0.7$ (blue lines)
  BHs. Solid and dashed lines represent the disc and jet components,
  respectively. The jet component was calculated assuming magnetic
  field saturation at half the MAD limit. $\dot M_{\rm ADAF}=10^{-3}\medd$
  is an estimated transition between optically thin and thick
  accretion flows.}
\label{f.Lfb}
\end{figure}

Fig.~\ref{f.Lfb} shows the
disc and jet components of the total black hole feedback, $L_{\rm fb}$
(Eqs.~\ref{e.Lfbthin} \& \ref{e.Lfbthick}), as a function of accretion
rate for BH spins $a_*=0.0$ and $0.7$. For the latter,  we assumed
that magnetic field saturated
at the BH at half of the MAD limit (in this way the jet power was not
overwhelming the power of the disc component). 

The solid lines show the
efficiency of the feedback coming from the disc. In the thin disc
regime ($\dot M_{\rm ADAF}\lesssim\dot M\lesssim\medd$), this
efficiency equals the standard thin disc efficiency - $\eta=0.057$ and
$0.104$ for $a_*=0.0$ and $0.7$, respectively. For accretion rates
significantly exceeding the Eddington limit, this efficiency drops
down to roughly half of the thin disc efficiency, i.e., to $\eta=0.03$
for a non-rotating BH. The proposed formulae make the transition
between the two regimes smooth. In the limit of low accretion rates
$\dot M<\dot M_{\rm ADAF}$, one expect accretion flows to be optically
thin with similar efficiency of $\eta=0.03$. The transition to the
thin disc limit is probably more violent, and we did not apply any
smoothening function there.

The dashed lines reflect the power of the jet feedback component. It
is non-zero only for the case of a rotating BH. For geometrically
thick discs jet production is efficient and given (for magnetic flux
at the BH saturated at half of the MAD limit)
by $1/4P_{\rm
  BZ}$ (see Eq.~\ref{e.PBZ}). Thin discs are unlikely to
provide strong jet components and therefore we damp the jet
power in this regime. One has to keep in mind that the jet component
will be highly collimated and may not interact efficiently with the ISM.

\section{Summary}
\label{s.summary}

In this paper we have studied the flow of energy in geometrically thick
discs, both optically thin and thick. We based our study on a set of
state-of-the-art, three-dimensional simulations of accretion flows
performed in the framework of general relativity. Our results are as
follows:

\begin{enumerate}
\item \textit{Total feedback:} Thick accretion flows on a non-rotating
  BH show the same total efficiency of $3\% \dot M c^2$ (roughly
  $50\%$ of the thin disc efficiency) independent
  of the accretion rate. Both optically thin, ADAF-like flows, and
  super-Eddington  optically thick flows liberate energy at the same
  rate. This energy is ultimately distributed between the energy
  carried by outflow and radiation.
The efficiency of accretion flows onto a  rotating BH is increased by the
modified spacetime geometry and the
rate at which BH rotational energy is extracted through the
Blandford-Znajek process. 

\item \textit{Approximated formulae:} One may
approximate the total amount
of feedback coming from an accreting system using
Eqs.~\ref{e.Lfbthin} and \ref{e.Lfbthick}. These formulae assume that
the total efficiency of the feedback disc component, i.e., the amount
of energy extracted from the disc itself, and not from the jet, equals half of the thin disc
efficiency for
geometrically thick discs, as found in this work. 

\item \textit{Energy in the outflow:} The energy outflowing from the system can be ultimately carried away
only by radiation or outflowing gas. If a disc cannot cool
efficiently, i.e., if it is not luminous (in radiation), then most of
the liberated energy must be carried by the outflow. This is
true both for optically thin discs and for optically thick discs
which at sufficiently high accretion rates efficiently trap radiation
in the gas. Therefore, we can
infer the existence of outflows even if they are not emerging strongly
within
the computational domain.

The amount of energy, either kinetic or
thermal, carried by the outflow equals,
\be
L_{\rm outflow}=L_{\rm fb}- L_{\rm rad},
\ee
where the radiative luminosity is zero for optically thin discs and
may be approximated as,
\be
L_{\rm rad}\approx L_{\rm Edd}\left(1+\log \dot M/\medd\right),
\ee
for super-critical discs.

\item \textit{Outflowing mass:} Our study is based on simulations
  covering only the innermost region of BH accretion. We find that
  significant amount of energy flows out from that region and likely
  results in outflowing mass from larger radii. However,
  despite the fact that we know how energetic the outflow can be, we
  are not able to say in what amount gas is blown away. The
  relation between the two depends on the Bernoulli function of the
  outflowing gas, e.g., marginally bound gas will carry
  virtually zero energy per unit mass. Because of similar reasons we
  cannot determine the amount of momentum carried with the outflow. As \cite{begelman-12} points
  out, accretion through thick, advective discs leads to either winds
  or breezes. To find how much gas is lost on
  the way towards the BH, one has to solve the problem consistently on
  larger scales than covered by the simulations presented
  here. Recently, a significant progress in this direction has been
  made by \cite{yuan+15}  and \cite{bu+15}, who studied optically thin
  accretion flows and found that gas is likely
  lost between $r\approx40$ and the Bondi radius, and that the mass
  loss rate
  in the wind increases proportionally to radius according to $\dot
  M_{\rm out}=\dot M_{\rm BH} (r/40)$.

\item \textit{Angular distribution of feedback:} In the case of optically thin accretion, the liberated energy can
flow out in two channels. The jet component related to the extraction
of BH rotational energy is collimated along the axis and ultimately
results in a narrow, relativistic magnetized jet. The accretion
component flows outward in the bulk of the disc and is responsible for
driving the outflows at large radii or ultimately leaves the system in
convective eddies. Such energy flows will have a
quasi-spherical distribution in space and will likely interact
efficiently with ISM.

To some extent similar properties characterize outflows in case of
super-critical, optically thick discs. The jet component is likely to
be colimated along the axis, while the outflow component covers wide
range of angles. The radiation coming out of the system may have
initially a mildly collimated component in the funnel region \citep[the
radiative jet, see][]{sadowski+radjets}. However, it either converts
into kinetic jet (if there is enough coupling between radiation and
gas in the funnel), or ultimately diffuses when the funnel opens because of only mild
collimation of the photon beams \citep{jiang+14b,narayan+heroic}. Therefore, the radiation component
should be expected to cover large solid angle from the point of view of a
distant observer.
Thin accretion discs, which we did not study here, are expected to
produce largely isotropic radiative feedback.

\item \textit{Models of thick discs:} Our study shows that
    outflows in some form or convection is inevitable for thick discs. This is not surprising because
    advection dominated accretion involve fluid which is only weakly
    bound to the BH \citep{narayanyi-94,adios}. Existence of outflows
    or convection in principle rules out well
  known and celebrated models of thick accretion flows which assume
  that gas is not lost on the way towards the BH, and which do not
  allow for convection, i.e., optically thick slim discs
  \citep{abramowicz+88}  and optically thin ADAFs
  \citep{narayanyi-94,abramowicz+adafs}. However, the outflow and
  convective regions
  do not extend all the way down to the BH. Therefore, the innermost
  region \textit{can} be described with the use of these models. 

In the outer region the situation is less clear because all the
proposed semi-analytic models for convection and winds suffer
from some problems, or they have been developed, as the recent
inflow-outflow solution by \cite{begelman-12}, in application for
the inner part of the flow. In particular, the models of convectively
dominated discs \citep{quataert-cdafs,narayan-cdafs,narayan-cdafs2,abramowicz-cdafs}
in optically thin flows are self-similar. The \cite{dotan+11}  model for slim discs with winds uses sophisticated
descriptions of the disc and the wind separately, but assumes
an ad hoc wind launching mechanism. Finally, simple and widely
used ADIOS model \citep{adios} takes strong assumptions which have been
criticized by \cite{abramowicz+00}.

\end{enumerate}

\section{Acknowledgements}

AS acknowledges support
for this work 
by NASA through Einstein Postdoctoral Fellowship number PF4-150126
awarded by the Chandra X-ray Center, which is operated by the
Smithsonian
Astrophysical Observatory for NASA under contract NAS8-03060. AS thanks
Harvard-Smithsonian Center for Astrophysics for its hospitality.
This research was supported by the Polish NCN grants UMO-2013/08/A/ST9/00795 and DEC-2012/04/A/ST9/00083.
JPL was supported in part by a grant from the French Space Agency CNES.
RN was
supported in part by NSF grant AST1312651 and NASA grant TCAN
NNX14AB47G.
The authors acknowledge computational support from NSF via XSEDE resources
(grant TG-AST080026N), and
from NASA via the High-End Computing (HEC) Program
through the NASA Advanced Supercomputing (NAS) Division at Ames
Research Center.
 
\bibliographystyle{mn2e}

\end{document}